\documentclass[final,5p,times,twocolumn]{elsarticle}

\usepackage{graphicx} 
\usepackage[framemethod=TikZ]{mdframed}
\usepackage{tikz}
\usepackage[x11names]{xcolor}
\usepackage[most]{tcolorbox}
\usepackage{lipsum}
\usepackage{cuted} 
\usepackage{xfp} 

\usepackage{mathtools}
\usepackage{amsthm}
\usepackage{thmtools}
\usepackage{amsmath}
\usepackage{amssymb}
\usepackage{cancel}
\usepackage{braket}
\usepackage{mathrsfs}
\usepackage{slashed}
\usepackage{comment}
\usepackage{bm}
\usepackage{etoolbox}
\usepackage{dsfont} 

\usepackage[T1]{fontenc}

\usepackage[colorlinks=true, linkcolor=blue, citecolor=blue, filecolor=blue, urlcolor=black]{hyperref} 
\usepackage{caption}
\usepackage{cleveref}

\biboptions{numbers,sort&compress}

\journal{Physics Letters B}

\begin{document}

\begin{frontmatter}

\title{Energy loss predicts no \texorpdfstring{$v_2$}{v2} in small systems}

\author[label1,corref{cor1}]{Ben Bert}
\author[label1]{Coleridge Faraday}
\author[label1,label2,label3]{W.A. Horowitz}

\affiliation[label1]{organization={University of Cape Town, Department of Physics, Private Bag X3},
            addressline={Rondebosch}, 
            city={Cape Town},
            postcode={7701}, 
            state={Western Cape},
            country={South Africa}}

\affiliation[label2]{organization={Okinawa Institute of Science and Technology Graduate University},
            addressline={1919-1 Tancha, Onna-son},
            city={Kunigami-gun},
            postcode={904-0495},
            state={Okinawa},
            country={Japan}}

\affiliation[label3]{organization={Department of Physics, New Mexico State University},
            city={Las Cruces},
            postcode={88003},
            state={New Mexico},
            country={USA}}

\cortext[cor1]{Corresponding author: Ben Bert, brtben004@myuct.ac.za}

\begin{abstract}
We present high-$p_T$ $R_{AB}$ and $v_2$ from a perturbative quantum chromodynamics-based energy loss model that includes event-by-event hydrodynamic evolution of the medium and small system size corrections to the energy loss. The model is calibrated on, and describes well, large system $R_{AA}$ and $v_2$ experimental data. The extrapolation of our model to $\mathrm{Ne}+\mathrm{Ne}$ and $\mathrm{O}+\mathrm{O}$ agrees quantitatively with recent experimental measurements of $R_{AA}$. Surprisingly, at high-$p_T$ our energy loss model predicts $v_2\approx0$ for all symmetric and asymmetric small systems when extracted using either hard-hard or hard-soft two-particle correlations. We argue that \emph{all} energy loss models will in general predict $v_2\approx0$ when extracted using hard-soft correlations, which is the usual experimental method for measuring anisotropy in hadronic collisions, due to a generic geometric decorrelation between the hard and soft sector participant planes.
\end{abstract}

\end{frontmatter}

\section{Introduction}
\label{Introduction}
There is overwhelming experimental evidence for the formation of the quark-gluon plasma (QGP) in central heavy-ion collisions in the large collision systems of Au+Au and $\mathrm{Pb}+\mathrm{Pb}$ at RHIC and the LHC, respectively \cite{Harris:1996zx,Niida:2021wut,Harris:2023tti,Pasechnik:2016wkt}. Some of the most notable evidence is seen in large low-$p_T$ $v_2$ \cite{STAR:2000ekf,STAR:2002hbo,ALICE:2010suc}, strangeness enhancement \cite{Rafelski:1982pu,STAR:2003jis,ALICE:2013xmt}, and quarkonium suppression \cite{Matsui:1986dk,CMS:2012bms,ALICE:2012jsl}. Similar success has been achieved in the hard sector, where Bjorken's original proposal of medium induced jet quenching \cite{Bjorken1982} is qualitatively supported by the nuclear modification factor $R_{AB}\equiv (d N^{A B} / d p_T) /(\left\langle N_{\text {coll }}\right\rangle d N^{p p } / d p_T) \ll1$  for high-$p_T$ (>10 GeV) hadrons \cite{ALICE:2010yje,CMS:2012aa,ATLAS:2015qmb,CMS:2016xef}. 

In the high-$p_T$ sector, while the azimuthally averaged $R_{AA}(p_T)$ has long been reproducible within a wide class of models \cite{Vitev:2002pf,Gyulassy:2004zy,Wang:2025lct}, achieving a simultaneous description of both azimuthally averaged high-$p_T$ $R_{AA}(p_T)$ and the angular dependence of $R_{AA}(p_T,\phi)$ captured by high-$p_T$ $v_2$ posed a significant challenge, as early calculations that described $R_{AA}$ typically under predicted $v_2$ \cite{Betz:2014cza,Xu:2014tda}. This tension---often referred to as the $R_{AA}\otimes v_2$ \textit{puzzle}---has been relieved in large systems through the development of more sophisticated energy loss frameworks that incorporate realistic medium evolution and event-by-event geometry fluctuations \cite{Noronha-Hostler:2016eow,Zigic:2018ovr,Xu:2015bbz}. Thus, the simultaneous description of high-$p_T$ $R_{AA}$ and $v_2$ in large systems is consistent with the interpretation of path-length dependent energy loss of hard partons traversing the QGP. Whether this interpretation extends to small systems is the central question of this study.

For the asymmetric small systems of $p/d+\mathrm{Pb}/\mathrm{Au}$, measurements show that the same observables---large low-$p_T$ $v_2$ \cite{ATLAS:2012cix,ATLAS:2013jmi,ALICE:2014dwt,CMS:2015yux}, strangeness enhancement \cite{ALICE:2013wgn,ALICE:2015mpp}, and quarkonium suppression \cite{ALICE:2016sdt}---are all qualitatively consistent with the formation of small droplets of QGP. However, in the hard sector, the message is not as clear: measurements of central high-$p_T$ $R_{p \mathrm{Pb}}\sim 1.2$ by ATLAS \cite{ATLAS:2022kqu} and $R_{p \mathrm{Pb}}\sim 1$ by ALICE \cite{ALICE:2021est} are inconsistent with the $R_{p \mathrm{Pb}}<1$ prediction of final state energy loss \cite{FH2025-2,Zhang:2013oca,Kang:2015mta,Ke:2022gkq}; in stark contrast, the PHENIX collaboration \cite{PHENIX:2023dxl} has measured a photon normalized $R_{d\mathrm{Au}}\sim0.75$ that is in qualitative agreement with the energy loss picture \cite{FH2025-2,Zhang:2013oca,Kang:2015mta,Ke:2022gkq}.

Non-zero and positive high-$p_T$ $v_2$ measurements for the small collision system of $p+\mathrm{Pb}$ have been reported by the ATLAS \cite{ATLAS:2019vcm}, ALICE \cite{ALICE:2022cwa} and CMS \cite{CMS:2025kzg} collaborations; these results are suggestive of path length dependent partonic energy loss in small systems. However, as mentioned, there is no clear suppression of $R_{p \mathrm{Pb}}$ below unity. This complicates the interpretation of the observed $v_2$ as an energy loss effect, since a path-length-dependent energy loss mechanism generically produces anisotropy \emph{and} suppression in tandem. The recent discovery of suppression of charged-particle production in $\mathrm{O}+\mathrm{O}$ collisions \cite{CMS:2025bta, ALICE:2025oop} may thus provide a valuable opportunity to study path-length dependent energy loss through a high-$p_T$ $v_2$ measurement in a small system where the signs of suppression induced by energy loss have been observed \cite{vanderSchee:2025hoe,Pablos:2025cli,Zakharov:2025mbk,Mazeliauskas:2025clt,FBBVW}. However, to establish whether an observed high-$p_T$ $v_2$ in small systems can be attributed to path-length dependent energy loss, it is crucial to have theoretical input that quantifies the expected high-$p_T$ $v_2$ from energy loss in small systems. 

The purpose of this Letter is to provide this quantitative theoretical input.

\section{Energy loss model}
\label{EnergyLossModelSection}
\begin{table}[t!]
    \centering
    \caption{Experimental datasets used in the global extraction of the effective strong coupling $\alpha_s$. All extractions use data from $\mathrm{Pb}+\mathrm{Pb}$ collisions at $\sqrt{s_{NN}}=5.02$ TeV in $p_T$ ranges of $10~\mathrm{GeV} \leq p_T \leq 50~\mathrm{GeV}$ and in the centrality range of $0-50\%$.}
    \label{table:experimentalDataSetsForExtraction}
    \begin{tabular}{c @{\hspace{1.25cm}} c @{\hspace{1.25cm}} c}
\hline\hline
Experiment & Hadron species & Observable \\ \hline \hline \\[0 em]
ALICE \cite{ALICE:2018lyv} & $D^0$ & $R_{AA}$ \\

ALICE \cite{ALICE:2018vuu} & $h^{\pm}$ & $R_{AA}$ \\

ALICE \cite{ALICE:2019hno} & $\pi^{\pm}$ & $R_{AA}$ \\

ATLAS \cite{ATLAS:2022kqu} & $h^{\pm}$ & $R_{AA}$ \\

CMS \cite{CMS:2016xef} & $h^{\pm}$ & $R_{AA}$ \\

CMS \cite{CMS:2017qjw} & $D^0$ & $R_{AA}$ \\

ALICE \cite{ALICE:2017pbx} & $D$ & $v_2$ \\

ALICE \cite{ALICE:2018rtz} & $h^{\pm}$ & $v_2$ \\

ATLAS \cite{ATLAS:2018ezv} & $h^{\pm}$ & $v_2$ \\

CMS \cite{CMS:2017vhp} & $D^0$ & $v_2$ \\

CMS \cite{CMS:2017xgk} & $h^{\pm}$ & $v_2$ \\

CMS \cite{CMS:2020bnz} & $D^0$ & $v_2$ \\

CMS \cite{CMS:2022vfn} & $D^0$ & $v_2$ \\ \\ [0 em]
\hline\hline
\end{tabular}
\end{table}

The energy loss model we present in this work is a qualitative improvement of \cite{FH2025-3}, which itself is based on the Wicks-Horowitz-Djordjevic-Gyulassy (WHDG) convolved radiative and collisional energy loss formalism \cite{WHDG2007}. The radiative energy loss formalism is an extension of the Djordjevic-Gyulassy-Levi-Vitev (DGLV) opacity expansion \cite{DGLV1,DGLV2,DGLV3}, with short path length corrections developed in \cite{Kolbe:MSc,Kolbe:SPLC}. The collisional energy loss of the model presented in this manuscript is based on Hard Thermal Loops (HTL) effective field theory \cite{HTL1,HTL5,HTL2,HTL3,HTL4,BT1,BT2}, and is calculated from two approaches. The first (BT) is based on the work of Braaten and Thoma \cite{BT1,BT2} and makes use of HTL propagators at small momentum transfers and vacuum propagators at large momentum transfers. The second approach we include (HTL-only) is based on the work of \cite{WicksPHD} and is calculated with HTL propagators for all momentum transfers.

In this work we improve on the treatment of the geometry in the energy loss compared to previous works, which used the static brick energy loss approximation \cite{FBBVW,FH2023,FH2025-1,FH2025-2,FH2025-3,FH2023SAIP}.
Here we dynamically model the scattering centers $\bar{\rho}$ of the medium with a power-law, as motivated in \cite{Andres:2023jao}, \textit{via} the following form 
\begin{equation}
  \bar{\rho}= \bar{\rho}_0\left(\frac{\tau_0}{z}\right)^{\beta} \theta\left(t_c-z\right) \theta\left(z-\tau_0\right),
\end{equation}
where $\bar{\rho}_0$ parametrizes the initial density of scattering centers, $\tau_0$ is the formation time of the medium, $t_c$ is the time at which the medium hadronizes, and $\beta=1.2$\footnote{We choose the $\beta$ parameter by using the Levenberg-Marquardt method \cite{Weisstein2025LevenbergMarquardt} to fit a power-law of the form $1/z^{\beta}$ to the temperature profiles of a set of paths through central $\mathrm{Pb}+\mathrm{Pb}$ collisions.}. Due to the $\theta\left(z-\tau_0\right)$ and $\theta\left(t_c-z\right)$ terms, no contributions from pre-thermalization and post-hadronization times are included in the energy loss. Energy loss occurring during pre-thermalization and post-hadronization times is not well understood and is often neglected \cite{Wang:2003aw,Elfner:2020men,Ipp:2020nfu,Andres:2022bql,Avramescu:2023qvv,Barata:2024xwy,Pablos:2025cli}; investigating the sensitivity of our model to pre-thermalization and post-hadronization energy loss is the subject of future work.

We dynamically model the scattering centers $\bar{\rho}$ of the medium as undergoing relativistic viscous hydrodynamic evolution by fitting the $\bar{\rho}_0$ and $t_c$ parameters to a trajectory through the medium \textit{via} the relation $\bar{\rho}\propto T^3$ where $T$ is the temperature of the medium. The temperature profiles obtained from \cite{ShenPrivateComm} are generated through hydrodynamic simulations performed with initial conditions obtained from the IP-Glasma model \cite{IP-Glasma1,IP-Glasma2,IP-Glasma3}; the initial conditions are then evolved with the MUSIC viscous relativistic $(2+1)$D hydrodynamics code \cite{MUSIC1,MUSIC2,MUSIC3}. Partons are produced isotropically and are weighted according to the number of binary collisions ($N_{\text{coll}}$) at their initial position. The $\bar{\rho}_0$ and $t_c$ parameters are fit to the temperature of the medium on a path-by-path basis. This enables a realistic event-by-event calculation of the energy loss. In doing so, we find a numerical speed-up of approximately seven orders of magnitude compared to a calculation that samples the medium's evolution directly from the hydrodynamic temperature profiles; this speed up enables us to incorporate the medium's event-by-event evolution along the full parton trajectory within the DGLV energy loss formalism. The details of the scattering center modeling procedure will be presented in a future long paper \cite{BertFaradayHorowitz2026}. 

It should be noted that our choice of hydrodynamic model is not unique, and we could have used a different hydrodynamic model for our medium background, \textit{e.g.} \cite{Moreland:2014oya,Werner:2010aa,Shen:2014vra}. However, since all reasonable hydrodynamic models describe the low-$p_T$ data comparably well, we expect our results to be relatively insensitive to the specific choice of hydrodynamic model.

The implementation of the energy loss model  we have described above leaves the strong coupling constant $\alpha_s$ as an effective free parameter; we constrain the effective strong coupling $\alpha_s$ to large-system $R_{AA}$ and $v_2$ experimental data. The details of the fitting procedure are given in \cite{FH2025-3}. In summary, we perform a global $\chi^2$ fit to the $R_{AA}$ and $v_2$ data in $\mathrm{Pb}+\mathrm{Pb}$ collisions, where we include experimental data in the $p_T$ ranges of $10~\mathrm{GeV} \leq p_T \leq 50~\mathrm{GeV}$ and in the centrality range of $0-50\%$. We restrict our analysis to this $p_T$ and centrality range because, in this region of phase space, QGP formation is well established and the interpretation of suppression is not subject to significant selection biases or additional model dependencies \cite{ALICE:2022wpn,CMS:2024krd,PHENIX:2004vcz,STAR:2005gfr,ALICE:2018ekf}. The extraction process incorporates three types of uncertainties:  \textit{Type A} uncertainties, which include statistical and systematic components that are uncorrelated in $p_T$; \textit{Type B} uncertainties, which correspond to systematic effects that are correlated in $p_T$ but whose detailed correlation structure is unknown; and \textit{Type C} uncertainties, which are fully correlated in $p_T$ and are dominated by overall normalization effects. \Cref{table:experimentalDataSetsForExtraction} lists the \protect\input{dataForCouplingExtraction/numberOfExperimentalInPointsGlobalFitDATApT1050LHCRAAv2.text}LHC data points used in our global extraction of our effective strong coupling constant. In this work we focus on $R_{AA}$ and $v_2$ measurements from the LHC at $\sqrt{s_{NN}}=5.02~\mathrm{TeV}$ and thus only include LHC experimental data from collisions in our effective coupling $\alpha_s$ extractions; in future work we plan to extend our analysis to include RHIC data.

\begin{figure*}[t!]
  \centering
  \begin{minipage}{0.71\linewidth}
    \centering
    \includegraphics[width=0.475\linewidth]{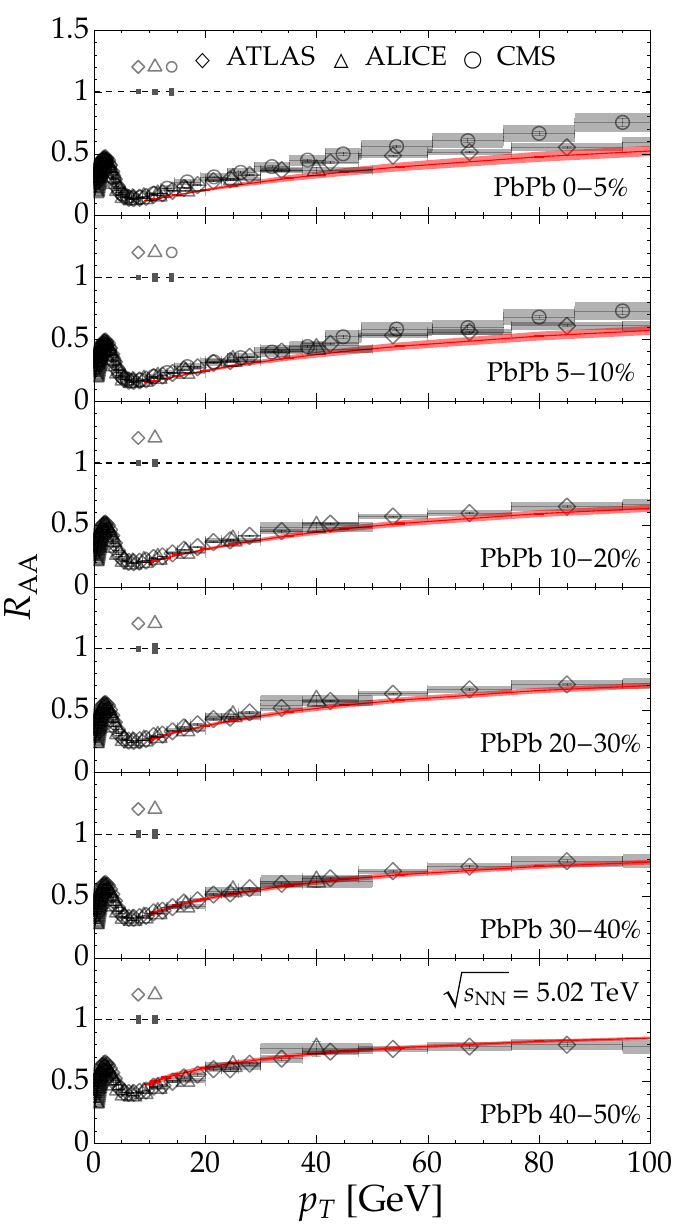}
    \includegraphics[width=0.4875\linewidth]{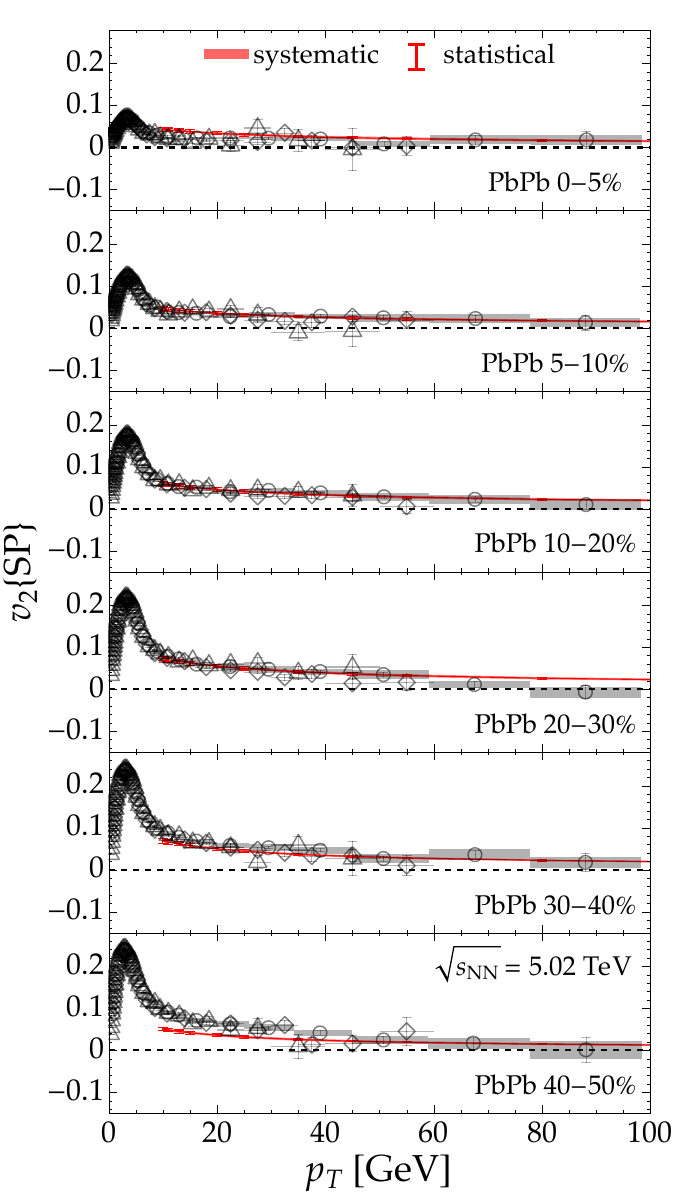}
  \end{minipage}
  \begin{minipage}{0.274\linewidth}
    \centering
    \includegraphics[width=\linewidth,
    trim=0mm 0mm 0mm 0mm,clip]{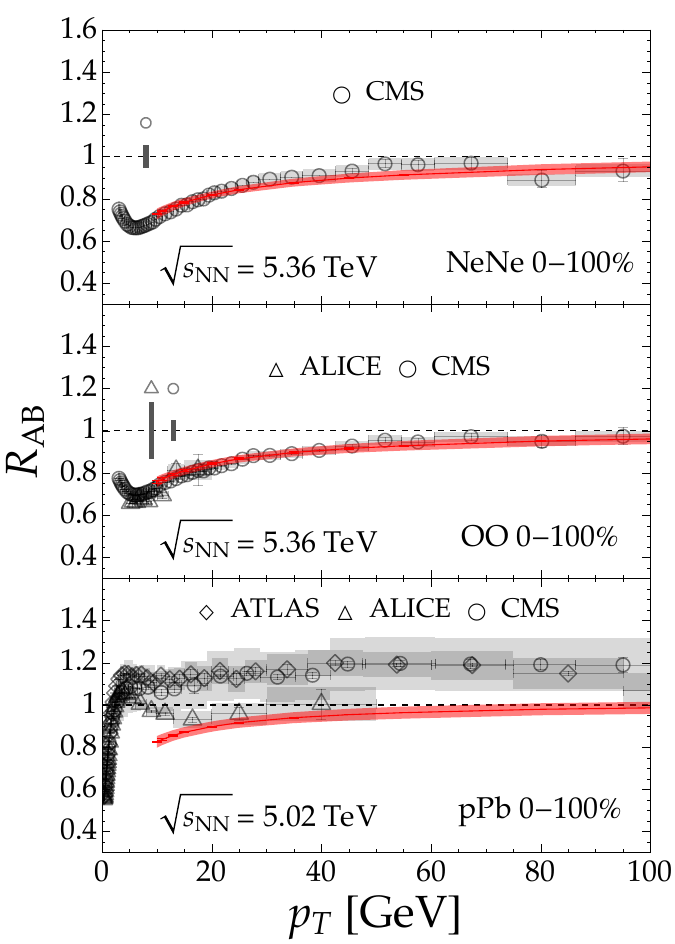}\\[0.5em]
    \includegraphics[width=\linewidth,
    trim=6mm 4mm 3mm 0mm,clip]{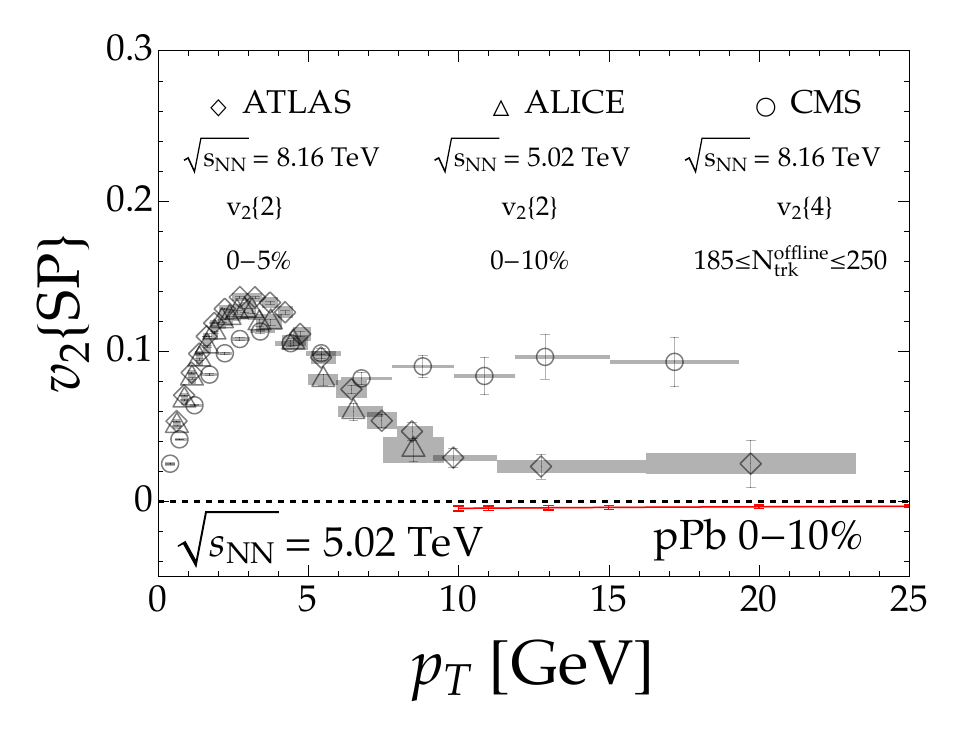}
  \end{minipage}  
  \caption{The left and middle columns compare energy loss model predictions for $\mathrm{Pb}+\mathrm{Pb}$ $R_{AA}$ and $v_2\{\text{SP}\}$ at $\sqrt{s_{NN}}=5.02$ TeV, respectively, as functions of $p_T$ across multiple centrality classes, with data from ATLAS \cite{ATLAS:2022kqu,ATLAS:2018ezv}, ALICE \cite{ALICE:2018vuu,ALICE:2018rtz}, and CMS \cite{CMS:2016xef,CMS:2017xgk}. The right column presents comparisons to small-system data: the top panels show $R_{AB}$ for $\mathrm{Ne}+\mathrm{Ne}$ (first) at $\sqrt{s_{NN}}=5.36$ TeV \cite{CMS:2026qef}, $\mathrm{O}+\mathrm{O}$ (second) at $\sqrt{s_{NN}}=5.36$ TeV \cite{CMS:2025bta,ALICE:2025oop}, and for $p+\mathrm{Pb}$ (third) at $\sqrt{s_{NN}}=5.02$ TeV \cite{CMS:2016xef,ALICE:2018vuu,ATLAS:2022kqu}, while the bottom panel shows $v_2\{\text{SP}\}$ predictions for $p+\mathrm{Pb}$ compared to ATLAS \cite{ATLAS:2019vcm} (0-5\%, $\sqrt{s_{NN}}=8.16$ TeV, $v_2\{2\}$), ALICE \cite{ALICE:2022cwa} (0-10\%, $\sqrt{s_{NN}}=5.02$ TeV, $v_2\{2\}$), and CMS \cite{CMS:2025kzg} ($185 \leq N_{\text {trk }}^{\text {offline }}<250\approx 0-5\%$, $\sqrt{s_{NN}}=8.16$ TeV, $v_2\{4\}$, four-subevent method). Experimental data are shown with statistical (bars) and systematic (boxes) uncertainties, while theory results are shown in red with statistical (bars) and systematic (bands) uncertainties, along with central values (lines). Diamonds, triangles, and circles denote ATLAS, ALICE, and CMS data, respectively. Bands around unity are experimental normalization uncertainties. Normalization uncertainties for $R_{p\mathrm{Pb}}$ are $4.6\%,~ 3.9\%, \text{ and } 2.3\%$ for ATLAS \cite{ATLAS:2022kqu}, ALICE \cite{ALICE:2018vuu}, and CMS \cite{CMS:2016xef}, respectively; the $R_{p\mathrm{Pb}}$ uncertainties are not shown for visual purposes.}
  \label{RAAV2LargeSystems}
\end{figure*}

In this work, we will follow \cite{FH2025-3} and include systematic theoretical uncertainties from two sources within the energy loss model. The first source we consider is the upper limit $|\mathbf{k}|_{\rm max}$ on the transverse radiated gluon momentum $|\mathbf{k}|$. The upper bound on $|\mathbf{k}|$ was motivated to ensure that the collinear and large formation time approximations used in the derivation of the DGLV radiative energy loss remain valid. The effects of including this upper bound are discussed in detail in \cite{FH2025-3,Horowitz:2009eb}. In order to quantify the sensitivity of our model to the upper bound of $|\mathbf{k}|$, we vary $|\mathbf{k}|_{\rm max}$ by factors of 0.5 and 2 and call this factor the ``$|\mathbf{k}|_{\rm max}$ multiplier.'' The second source of theoretical uncertainty that we consider arises from the transition between HTL and vacuum propagators for collisional energy loss. We account for our model's sensitivity to the transition between HTL and vacuum propagators by calculating the collisional energy loss by using both the HTL-only and the BT formalism. We follow \cite{FH2025-3} and implement the theoretical uncertainty of our model's prediction by treating each choice of the $\mathbf{k}_{\rm max}$ multiplier and each choice of the collisional energy loss calculation (HTL-only vs BT) as an independent model and apply the statistical analysis mentioned above to each of these models. For each model, we extract an effective strong coupling constant from the data in \cref{table:experimentalDataSetsForExtraction}, and find values ranging from \protect\input{dataForCouplingExtraction/averageExtractedCouplingpT1050LHCRAAv2Range.text}for the effective coupling. Additionally, and independently of the systematic theoretical uncertainty, we include statistical theoretical uncertainty that arises from the finite number of events used in the calculation of the observables; the statistical theoretical uncertainty is calculated by performing a bootstrap analysis \cite{EfronTibshirani1993} on the events used in the calculation of the $R_{AB}$ and $v_2$ observables---see \cite{BertFaradayHorowitz2026} for details. In the figures in this Letter the systematic theoretical uncertainty of the model is shown as a band and the statistical theoretical uncertainty is shown with error bars. It is important to note that the systematic theoretical uncertainties considered here do not exhaust all possible sources of model uncertainty; \textit{e.g.}, we do not account for uncertainties associated with the choice of the onset time for energy loss \cite{Zigic:2019sth,Andres:2019eus}, energy loss occurring after hadronization \cite{Dorau:2019ozd,Andres:2019eus,Datta:2025gql}, or the inclusion of a running coupling \cite{Buzzatti:2012dy,Zakharov:2018cpv}. We leave a quantitative estimate of these uncertainties for future work.

In order to make contact with experimental data, we characterize the anisotropy through the angular nuclear modification factor $R_{AB}(p_T,\phi)$, which can be Fourier decomposed in the azimuthal angle $\phi$ as \cite{Noronha-Hostler:2016eow}
\begin{equation}
  \label{RAAFourierEquation}
  \frac
  {R_{A B}\left(p_T, \phi\right)}
  {R_{A B}\left(p_T\right)}
  =
  1+2 
  \sum_{n=1}^{\infty} 
  v_n^{\text {hard }}\left(p_T\right) 
  \cos 
  \left[
    n
    \left(
      \phi- \psi_n^{\text {hard }}\left(p_T\right)
    \right) 
  \right],
\end{equation}
where the $v_n^{\text {hard }}$ Fourier coefficients take the form
\begin{equation}
  \label{vnHardEq}
  v_n^{\text {hard }}\left(p_T\right) 
  =
\left\langle 
    \cos\left[{n\left(\phi-\psi_n^{\text {hard }}\left(p_T\right) \right)}\right]
  \right\rangle,
\end{equation}
with $\langle\cdots\rangle$ denoting an average over $\phi$ and events. The $\psi_n^{\text {hard }}$ reference angles from \cref{RAAFourierEquation} can be found through the following expression \cite{Noronha-Hostler:2016eow}
\begin{equation}
  \begin{aligned}
    \label{psi_nEQ}
    \psi_n^{\text {hard }}=
    \frac{1}{n}
    \text{arctan}2
    \Bigg(
      &\displaystyle{\int_0^{2 \pi}} d \phi~ R_{AB}\left(p_T, \phi\right) \cos (n \phi),\\
      &\displaystyle{\int_0^{2 \pi}} d \phi~ R_{AB}\left(p_T, \phi\right) \sin (n \phi)
    \Bigg).
  \end{aligned}
\end{equation}
In \cref{psi_nEQ}, $\text{arctan}2$ is the two-argument arctangent function that keeps track of the signs of the numerator and denominator to determine the correct quadrant of the angle \cite{IP-Glasma1}.

Anisotropy is commonly experimentally measured through the scalar-product (SP) method, $v_n\{\text{SP}\}$, and the two particle cumulant method, $v_n\{2\}$. The two measurements of anisotropy become identical, in principle, once non-flow contributions are subtracted from the two particle cumulant \cite{Luzum:2013yya,Snellings:2011sz}.  In this work we will report the $v_n^{ \text{hard}}$ anisotropy from \cref{vnHardEq} and the scalar-product anisotropy $v_n\{\text{SP}\}$, which is defined as \cite{CMS:2017xgk,ATLAS:2018ofq,ATLAS:2018ezv,ATLAS:2024mch,ALICE:2014qvj,CMS:2017vhp,CMS:2020bnz,CMS:2022vfn}
\begin{equation}
  \label{vnSPDefEQ}
  v_n\{\text{SP}\}(p_T)
  \equiv
  \frac
    {1}
    {N_h}
  \frac
    { 
    \sum_{\ell=1}^{N_e}
    \sum_{j=1}^{N_h}
    \vec{Q}_n^\ell\cdot\vec{u}_n^{\text{ }\ell j}(p_T)
    }
    {
    \sum_{\ell=1}^{N_e}
    \sqrt{
    \vec{Q}_n^\ell\cdot\vec{Q}_n^\ell
    }
    },
\end{equation}
where $\vec{u}_n^{\text{ }\ell j}\equiv(\cos n\phi^{\ell j},\sin n\phi^{\ell j})$ is constructed for each $p_T$ bin from the azimuthal angle $\phi^{\ell j}$ (taken relative to the beam axis) of the $j^{\text{th}}$ hadron in the $\ell^{\text{th}}$ event, and $\vec{Q}_n^\ell$—which approximates the soft participant plane—is obtained from the real and imaginary parts of $\mathcal{Q}_n^\ell\equiv\sum_k e^{in\phi^{\ell k}}$. Here $N_h$ and $N_e$ denote the number of hadrons per event and the total number of events, respectively. The $v_n\{\text{SP}\}$ can be shown to be related to the $v_n^{\text{hard}}$ \textit{via} the following relation:
\begin{equation}
  \label{vnSP_and_vnHard}
  v_n\{\text{SP}\}(p_T)
  =
  \frac
  {\left\langle \sqrt{\vec{Q}_n\cdot\vec{Q}_n} ~\right\rangle}
  {\sqrt{\left\langle \vec{Q}_n\cdot\vec{Q}_n \right\rangle}}
  \cos
  \left[
    n
    \left(
      \psi_n^{\text {hard }}(p_T)-\psi_n^{\text {soft }}
    \right)
  \right]
  v_n^{\text {hard }}(p_T),
\end{equation}
where $\langle\cdots\rangle$ now denotes averaging over events, and $\psi_n^{\text{hard}}$ and $\psi_n^{\text{soft}}$, defined in \cref{psi_nEQ} and \cref{psiNSoftEqDef}, align with the hard and soft participant planes, respectively. The soft-sector participant plane angles $\psi_n^{\text {soft }}$ are defined as 
\begin{equation}
  \label{psiNSoftEqDef}
  \psi_n^{\text {soft }}\equiv\frac{1}{n}\text{arctan}2\left(Q_n^x,Q_n^y\right).
\end{equation}

\begin{figure}[t!]
  \includegraphics[width=1\linewidth]{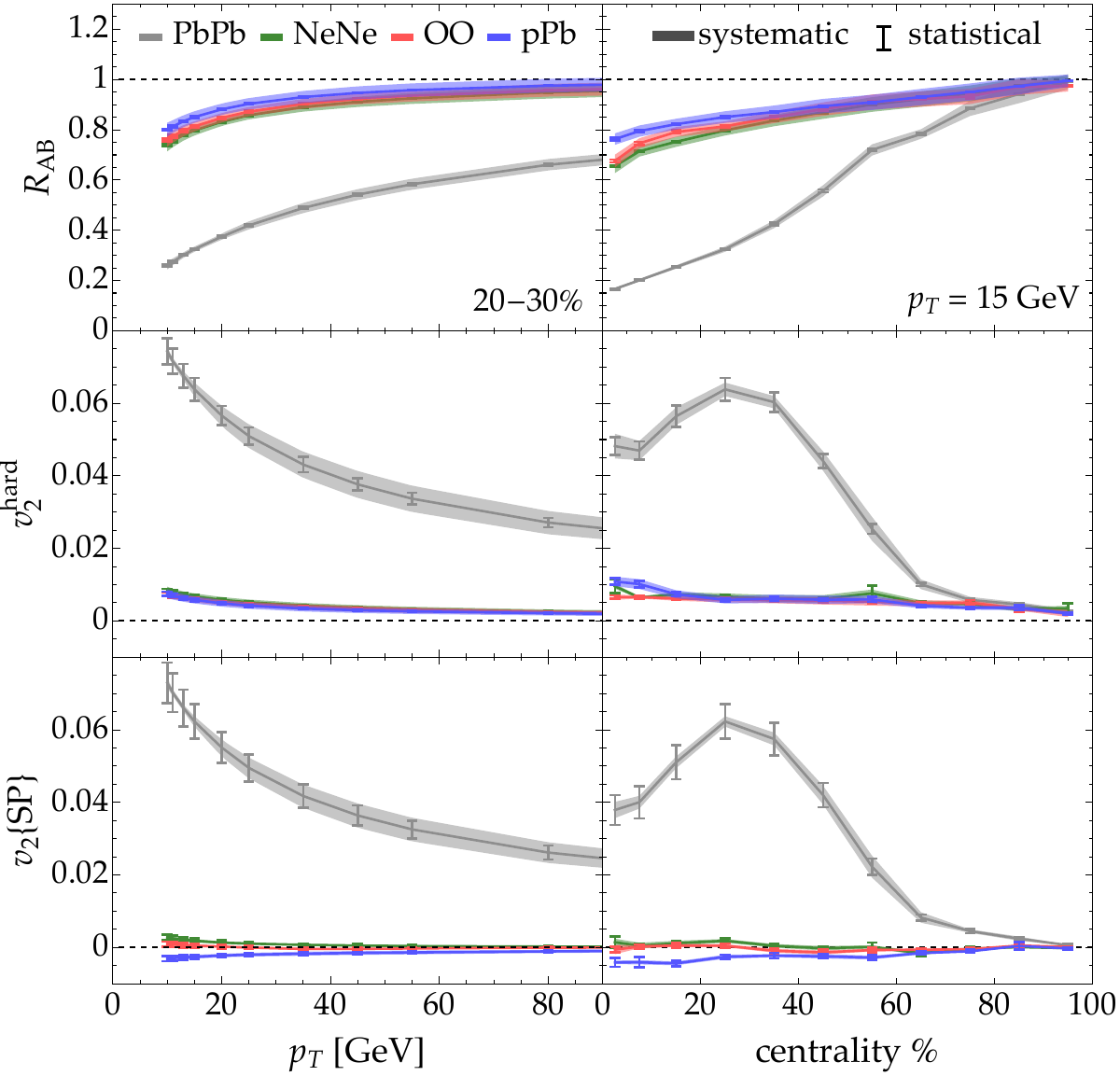}
  \caption{
Left (Right): Model predictions for $R_{AB}$ (top), $v_2^{\text{hard}}$ (middle), $v_2\{\text{SP}\}$ (bottom) as a function of $p_T$ (centrality) for $\mathrm{Pb}+\mathrm{Pb}$ (gray), $\mathrm{Ne}+\mathrm{Ne}$ (green), $\mathrm{O}+\mathrm{O}$ (red), and $p+\mathrm{Pb}$ (blue) collision systems. The left panel shows model predictions for the 20-30\% centrality class and the right panel shows model predictions at a fixed $p_T$ of 15 GeV. The $v_2^{\text{hard }}$ is calculated from \cref{vnHardEq} and the $v_2\{\text{SP}\}$ is calculated from \cref{vnSPDefEQ}. Results are shown with statistical uncertainties (bars), systematic uncertainties (bands), and central values (lines).
}
  \label{RAAv2SPV2Fourier}
\end{figure}

In the left and middle columns of \cref{RAAV2LargeSystems}, we compare the fitted results from the constrained energy loss model to representative large-system experimental data from ATLAS, ALICE, and CMS for $R_{AA}$ \cite{ATLAS:2022kqu,ALICE:2018vuu,CMS:2016xef} and $v_2$ \cite{ATLAS:2018ezv,ALICE:2018rtz,CMS:2017xgk} in $\mathrm{Pb}+\mathrm{Pb}$ collisions at $\sqrt{s_{NN}}=5.02~\mathrm{TeV}$. In the right column of \cref{RAAV2LargeSystems}, we present model predictions for small systems. The upper panels show $R_{AB}$ for $\mathrm{Ne}+\mathrm{Ne}$ \cite{CMS:2026qef} (first) and $\mathrm{O}+\mathrm{O}$ \cite{CMS:2025bta,ALICE:2025oop} (second) at $\sqrt{s_{NN}}=5.36$ TeV , and for $p+\mathrm{Pb}$ \cite{CMS:2016xef,ALICE:2018vuu,ATLAS:2022kqu} (third) at $\sqrt{s_{NN}}=5.02$ TeV. The lower panel shows $v_2\{\text{SP}\}$ predictions for $p+\mathrm{Pb}$ collisions at 0-10\% centrality with $\sqrt{s_{NN}}=5.02$ TeV and is compared to ATLAS \cite{ATLAS:2019vcm}, ALICE \cite{ALICE:2022cwa}, and CMS \cite{CMS:2025kzg} data. The ATLAS data corresponds to a measurement of $v_2\{2\}$ at 0-5\% centrality with $\sqrt{s_{NN}}=8.16$ TeV, the ALICE data corresponds to a measurement of $v_2\{2\}$ at 0-10\% centrality with $\sqrt{s_{NN}}=5.02$ TeV, and the CMS data corresponds to a measurement of $v_2\{4\}$ using the four-subevent method with $185 \leq N_{\text {trk }}^{\text {offline }}<250 \approx 0-5\%$ centrality at $\sqrt{s_{NN}}=8.16$ TeV.

From \cref{RAAV2LargeSystems} one can see that our model gives simultaneously a good quantitative description of the $p_T$ and centrality dependence of both the $R_{AA}$ and $v_2$ in large systems. Further, extrapolated to the small symmetric systems of $\mathrm{Ne}+\mathrm{Ne}$ and $\mathrm{O}+\mathrm{O}$, our model gives a good quantitative description of the minimum bias $R_{AA}$ as a function of $p_T$. As mentioned in the Introduction, the $R_{p\mathrm{Pb}}$ story is complicated; our model is in qualitative agreement with ALICE \cite{ALICE:2018vuu} but in qualitative disagreement with ATLAS \cite{ATLAS:2022kqu} and CMS \cite{CMS:2016xef} $R_{p\mathrm{Pb}}(p_T)$. On the other hand, our model predicts $v_2^{p \mathrm{Pb}}\{\text{SP}\}\approx 0$ in disagreement with data \cite{ATLAS:2019vcm,ALICE:2022cwa,CMS:2025kzg}.

This qualitative disagreement in the high-$p_T$ $v_2^{p \mathrm{Pb}}\{\text{SP}\}$ suggests that the energy loss model is significantly challenged by data. However, there are many potentially important theoretical and experimental complications associated with asymmetric collision systems \cite{ALICE:2014xsp,Alvioli:2013vk,Kordell:2016njg,Perepelitsa:2024eik,JETSCAPE:2024dgu}. In order to investigate the possible physical and experimental differences between symmetric and asymmetric collision systems we would thus like to predict the high-$p_T$ azimuthal anisotropy for the symmetric small systems $\mathrm{Ne}+\mathrm{Ne}$ and $\mathrm{O}+\mathrm{O}$ for comparison with future experimental analyses.

\begin{figure}[t!]
  \includegraphics[width=1\linewidth,
  trim=12mm 0mm 12mm 2.5mm,
  clip]{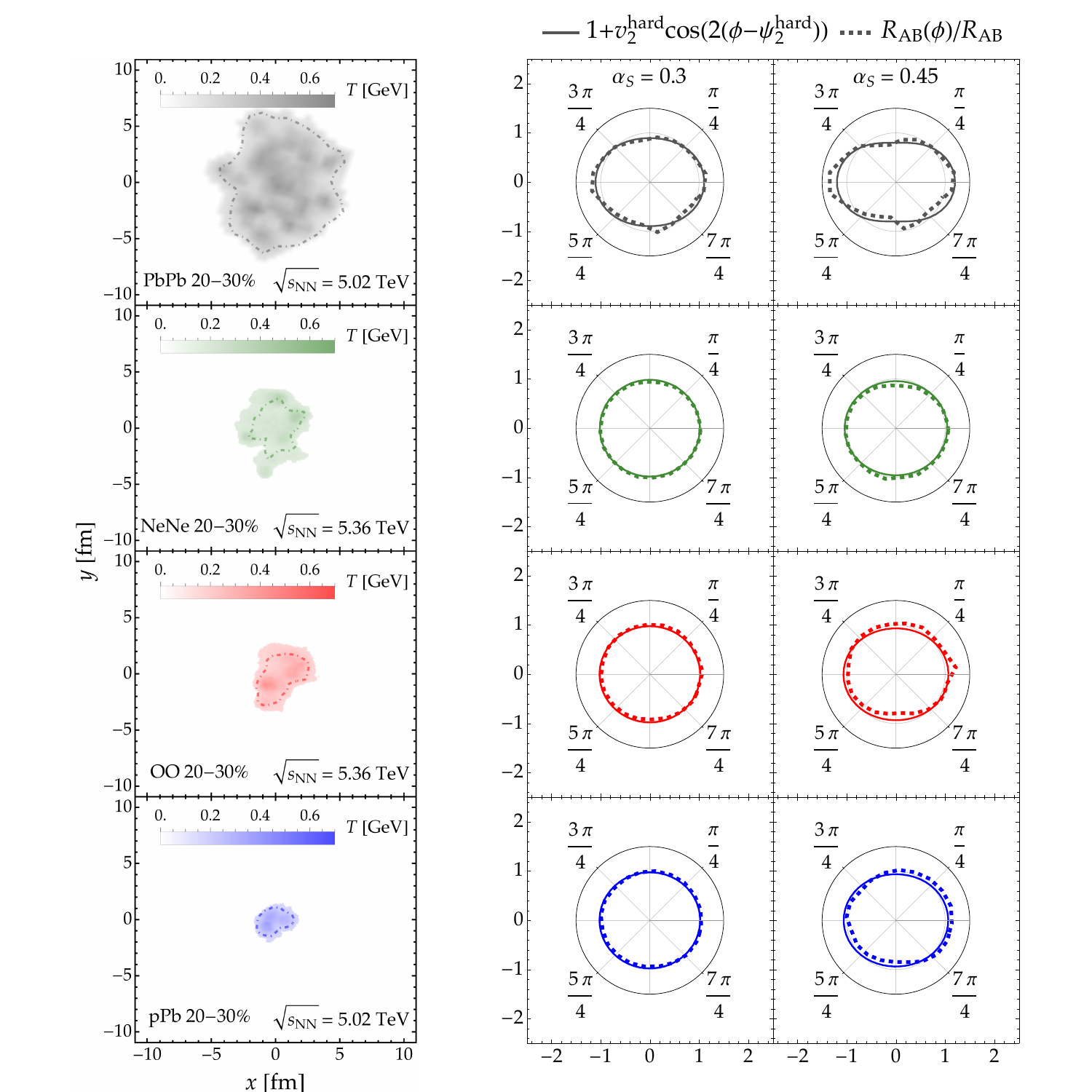}
  \caption{\protect\input{captions/hydroRAAEventPlotCaption.text}}
  \label{hydroRAAEventPlot}
\end{figure}

\section{Small system predictions for high-\texorpdfstring{$p_T$}{pT} \texorpdfstring{$R_{AB}$}{RAB} and \texorpdfstring{$v_2$}{v2}}
\label{resultsSection}

In \cref{RAAv2SPV2Fourier} we present the main results of this Letter: we show high-$p_T$ $R_{AB}$ (top row), $v_2^{\text{hard}}$ (middle row), and $v_2\{\text{SP}\}$ (bottom row) constrained model results for $\mathrm{Pb}+\mathrm{Pb}$ and predictions for the small systems of $\mathrm{Ne}+\mathrm{Ne}$, $\mathrm{O}+\mathrm{O}$, and $p+\mathrm{Pb}$. Results are shown as a function of $p_T$ (left column) and centrality (right column). In the top row we show that $R_{\mathrm{PbPb}}$ is significantly suppressed ($\sim 0.3$), while the small systems $R_{\mathrm{OO}}$, $R_{\mathrm{NeNe}}$ and $R_{p \mathrm{Pb}}$ all show similar signs of non-negligible suppression ($\sim 0.85$). Despite $R_{AB}$ predictions significantly less than unity, the $v_2^{\text{hard }}$ in all the small systems, $\mathrm{Ne}+\mathrm{Ne}$, $\mathrm{O}+\mathrm{O}$, and $p+\mathrm{Pb}$,  are small ($\sim0.005$) at $p_T\sim15$ GeV. In the bottom row of \cref{RAAv2SPV2Fourier} we show that while $v_2^{\text{hard}}\approx v_2\{\text{SP}\}$ for $\mathrm{Pb}+\mathrm{Pb}$, $v_2^{\text{hard}}>v_2\{\text{SP}\}\approx 0$ for the small systems of $\mathrm{Ne}+\mathrm{Ne}$, $\mathrm{O}+\mathrm{O}$, and $p+\mathrm{Pb}$.

The small $v_2^{\text{hard }}$ values in \cref{RAAv2SPV2Fourier} can be understood by examining the results of \cref{hydroRAAEventPlot}.
In the left column of \cref{hydroRAAEventPlot} we show hydrodynamic temperature profiles at the formation time of the medium ($\tau_0=0.4$ fm/c) of representative events for the collision systems of $\mathrm{Pb}+\mathrm{Pb}$ (top), $\mathrm{Ne}+\mathrm{Ne}$ (second), $\mathrm{O}+\mathrm{O}$ (third), and $p+\mathrm{Pb}$ (bottom). In the middle ($\alpha_s =0.3$) and right ($\alpha_s =0.45$) columns of \cref{hydroRAAEventPlot}, we compare $1+v_2^{\text{hard}}\cos\left(2(\phi-\psi^{\text{hard}}_2)\right)$ (solid) and $R_{A B}\left(p_T, \phi\right)/R_{A B}\left(p_T\right)$ (dotted). The results in the middle and right columns of \cref{hydroRAAEventPlot} are calculated using events which correspond to those in the left column of \cref{hydroRAAEventPlot}. The hydrodynamic temperature profiles have been rotated so as to align the $\psi_2^{\text{hard}}$ angles with the horizontal axis.

In \cref{hydroRAAEventPlot} one sees that there is an anisotropy in the hydrodynamic temperature profiles of the $\mathrm{Pb}+\mathrm{Pb}$, $\mathrm{Ne}+\mathrm{Ne}$, $\mathrm{O}+\mathrm{O}$, and $p+\mathrm{Pb}$ collision systems. The top row shows that our energy loss model is sensitive to the difference in path lengths that arise due to the anisotropy of the temperature profile in the large collision system of $\mathrm{Pb}+\mathrm{Pb}$. However, the bottom three rows show that despite increasing the strong coupling from 0.3 to 0.45, our energy loss model is weakly sensitive to the difference in path lengths that arise due to the anisotropy of the temperature profiles in the small collision systems of $\mathrm{Ne}+\mathrm{Ne}$, $\mathrm{O}+\mathrm{O}$, and $p+\mathrm{Pb}$.

\begin{figure}[t!]
  \includegraphics[width=0.94\linewidth]{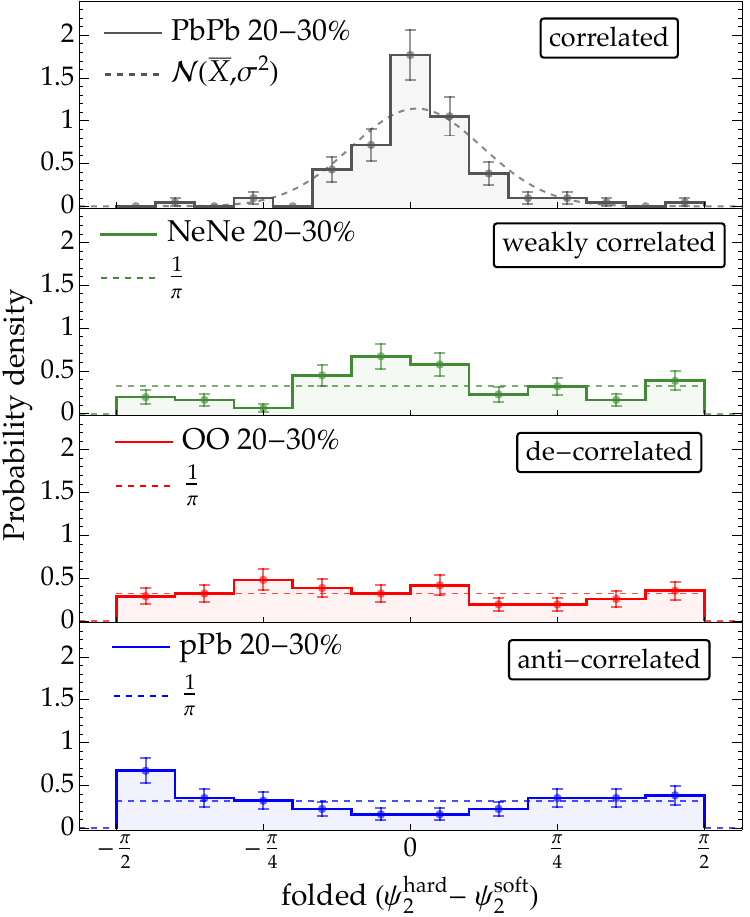}
  \caption{\protect\input{captions/decorrelationOfHardAndSoftReferenceAnglesPlotCaption.text}}
  \label{decorrelationOfHardAndSoftReferenceAnglesPlot}
\end{figure}

The $v_{2}\{\text{SP}\}\approx0$ result in small systems can be understood by examining \cref{decorrelationOfHardAndSoftReferenceAnglesPlot}, where we plot the folded distribution of $\psi_2^{\text {hard }}-\psi_2^{\text {soft }}$ for $\mathrm{Pb}+\mathrm{Pb}$ (top), $\mathrm{Ne}+\mathrm{Ne}$ (second), $\mathrm{O}+\mathrm{O}$ (third), and $p+\mathrm{Pb}$ (bottom) collision systems. The ``folded'' mapping maps all angles into the domain $[-\pi/2,\pi/2]$; this mapping is chosen to ensure that the magnitude of the largest difference between $\psi_2^{\text {hard }}$ and $\psi_2^{\text {soft }}$ occurs at $\pi/2$. 

We show in \cref{decorrelationOfHardAndSoftReferenceAnglesPlot} that the prediction of $v_{2}\{\text{SP}\}\approx0$ in small systems arises because the orientation of the participant plane defined by hard particles and the participant plane defined by soft particles become decorrelated. We see that for the $\mathrm{Pb}+\mathrm{Pb}$ collision system, the folded distribution of $\psi_2^{\text {hard }}-\psi_2^{\text {soft }}$ is peaked around zero and is well-approximated by a Gaussian of width $\sigma=0.35$ radians, and thus the participant planes in the hard and soft sectors are strongly correlated. Both $\mathrm{Ne}+\mathrm{Ne}$ and $\mathrm{O}+\mathrm{O}$ collisions systems are well-approximated by the uniform distribution $1/\pi$ over the entire angular range. Qualitatively, the $\mathrm{Ne}+\mathrm{Ne}$ system shows a slightly stronger correlation than the $\mathrm{O}+\mathrm{O}$ system, but the correlation is still weak compared to the $\mathrm{Pb}+\mathrm{Pb}$ system. Thus, we refer to the participant planes in the hard and soft sectors as weakly correlated and decorrelated in the $\mathrm{Ne}+\mathrm{Ne}$ and $\mathrm{O}+\mathrm{O}$ systems, respectively.
In the $p+\mathrm{Pb}$ system, the distribution of $\psi_2^{\text {hard }}-\psi_2^{\text {soft }}$ shows a double peak structure around $\pm \pi/2$, indicating that the participant planes in the hard and soft sectors are actually anti-correlated. From \cref{vnSP_and_vnHard}, one can see that the event averaged cosine when the hard- and soft-participant planes in small systems are decorrelated will result in $v_2\{\text{SP}\}\approx 0$. We will now argue why the $v_2\{\text{SP}\}\approx 0$ result is independent of the energy loss model used.

\begin{figure}[t!]
  \includegraphics[width=0.99\linewidth]{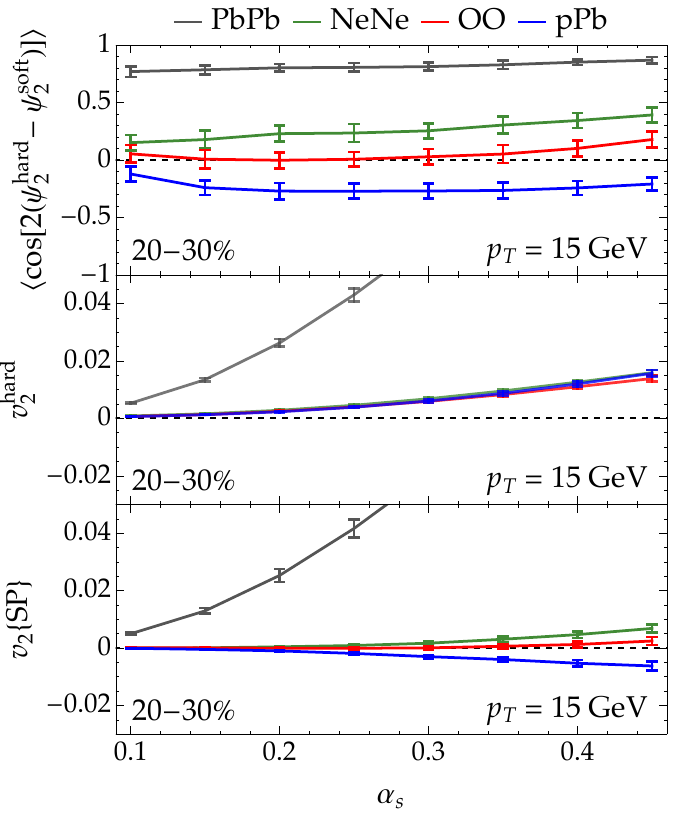}
  \caption{Dependence on the effective strong coupling $\alpha_s$ of the event-averaged $\cos(2[\psi_2^{\text{hard}}-\psi_2^{\text{soft}}])$ (top), $v_2^{\text{hard}}$ (middle), and $v_2\{\text{SP}\}$ (bottom) for $\mathrm{Pb}+\mathrm{Pb}$ (gray), $\mathrm{Ne}+\mathrm{Ne}$ (green), $\mathrm{O}+\mathrm{O}$ (red), and $p+\mathrm{Pb}$ (blue) collision systems. All results are shown for 20-30\% centrality at fixed $p_T=15$ GeV with $|\mathbf{k}|_{\rm max}$ multiplier set to 1. Statistical uncertainties are indicated by error bars.} 
  \label{correlationAndVsAgainstCouplingPlot}
\end{figure}

From \cref{vnSP_and_vnHard}, $v_2\{\text{SP}\}$ is determined by the flow vectors, the cosine of the difference between $\psi_2^{\text{soft}}$ and $\psi_2^{\text{hard}}$, and $v_2^{\text{hard}}$. The soft-sector flow vectors $Q_n$ and the reference angles $\psi_n^{\text{soft}}$ are determined entirely by the soft medium dynamics and are therefore energy loss model independent. While the magnitude of the $v_2^{\text{hard}}$ Fourier coefficients depends on the specific energy loss model used to calculate them, $v_2\{\text{SP}\}\approx 0$ will result so long as there is a decorrelation between $\psi_n^{\text{soft}}$ and $\psi_n^{\text{hard}}$. One might think that an energy loss model with a larger $v_2^{\text{hard}}$ would generate a greater correlation between $\psi_n^{\text{soft}}$ and $\psi_n^{\text{hard}}$. However, we show in \cref{correlationAndVsAgainstCouplingPlot} that the decorrelation between the event averaged cosine is largely insensitive to the magnitude of $v_2^{\text{hard}}$. Therefore, we expect that the prediction of $v_2\{\text{SP}\}\approx 0$ given by our model will be robust to changes across energy loss models.

In \cref{correlationAndVsAgainstCouplingPlot} we show the event event-averaged $\cos(2[\psi_2^{\text{hard}}-\psi_2^{\text{soft}}])$ (top), $v_2^{\text{hard}}$ (middle), and $v_2\{\text{SP}\}$ (bottom) as a function of $\alpha_s$ for $\mathrm{Pb}+\mathrm{Pb}$ (gray), $\mathrm{Ne}+\mathrm{Ne}$ (green), $\mathrm{O}+\mathrm{O}$ (red), and $p+\mathrm{Pb}$ (blue) collision systems at 20-30\% centrality with $p_T=15$ GeV.

One can see that as one scans across $\alpha_s$, even though the $v_2^{\text{hard}}$ remains relatively small as a function of $\alpha_s$, the significant relative change in $v_2^{\text{hard}}$ yields only a relatively small change in the event average of the cosine between the difference of $\psi_n^{\text{soft}}$ and $\psi_n^{\text{hard}}$. We take this insensitivity as confirmation that the decorrelation is robust against changes in energy loss models.

\section{Summary and outlook}

In this manuscript, we presented predictions for high-$p_T$ $R_{AB}$ and $v_2$ in $\mathrm{Ne}+\mathrm{Ne}$, $\mathrm{O}+\mathrm{O}$, and $p+\mathrm{Pb}$ collisions from a pQCD-based energy loss model. The energy loss model includes finite size corrections relevant for small systems, is constrained to large-system high-$p_T$ $R_{AA}$ and $v_2$ experimental data, incorporates event-by-event fluctuations in the soft sector, and calculates the high-$p_T$ $v_2$ using the same hard-soft correlation procedure employed experimentally. Within this framework, we demonstrated a simultaneous description of the high-$p_T$ $R_{AA}$ and $v_2$ data in central and semi-central $\mathrm{Pb}+\mathrm{Pb}$ collisions from the LHC.

Applying the model to small systems, we reproduced minimum bias $R_{AA}(p_T)$ for both $\mathrm{Ne}+\mathrm{Ne}$ and $\mathrm{O}+\mathrm{O}$ collision species. When comparing our model to minimum bias $p+\mathrm{Pb}$ data, we found qualitative agreement with ALICE $R_{p\mathrm{Pb}}(p_T)\lesssim 1$ \cite{ALICE:2018vuu} but were qualitatively inconsistent with ATLAS \cite{ATLAS:2022kqu} and CMS \cite{CMS:2016xef} $R_{p\mathrm{Pb}}(p_T)>1$. Our model resulted in $v_2\{\text{SP}\}\approx0$ for $p+\mathrm{Pb}$ which was in strong disagreement with measurements made by ATLAS \cite{ATLAS:2019vcm}, ALICE \cite{ALICE:2022cwa}, and CMS \cite{CMS:2025kzg}.

Our model predicted that for the small systems of $\mathrm{Ne}+\mathrm{Ne}$, $\mathrm{O}+\mathrm{O}$, and $p+\mathrm{Pb}$ for all centralities $v_2^{\text{hard}}\lesssim 0.01$ and $-0.005\lesssim v_2\{\text{SP}\}\lesssim 0.005$. We showed that our energy loss model was weakly sensitive to the anisotropy present in the initial temperature of the $\mathrm{Ne}+\mathrm{Ne}$, $\mathrm{O}+\mathrm{O}$, and $p+\mathrm{Pb}$ collision systems, and that the insensitivity resulted in a small $v_2^{\text{hard}}$. In Pb+Pb we saw that the $\psi_2^{\text {hard }}$ and $\psi_2^{\text {soft }}$ angles---which align the orientation of the anisotropy in the hard and soft sectors, respectively---were strongly correlated, but that these angles were approximately uncorrelated for all small systems considered here. Finally we showed that this decorrelation in small systems was approximately independent of the effective coupling in our model.

Since the decorrelation of hard and soft participant planes appears to be insensitive to the details of our energy loss model, we predict that $v_2\{\text{SP}\}\approx 0$ will generically hold across all energy loss models in all small symmetric and small asymmetric collision systems. If the $v_2\{\text{SP}\} >0$ measurement \cite{ATLAS:2019vcm,ALICE:2022cwa,CMS:2025kzg} in $p+\mathrm{Pb}$ collision systems is not due to energy loss, an obvious question is: what is the origin of the measured anisotropy? On the theory side it is difficult to imagine a simple parameter tuning to describe the large $v_2\{\text{SP}\}$ measurement; therefore, a theory-based resolution to the discrepancy likely requires additional physics.

One example of additional physics is sub-eikonal corrections \cite{Armesto:2004pt,Sadofyev:2021ohn,Fu:2022idl,Bahder:2024jpa}, but these corrections are generally small and become smaller with increasing $p_T$; thus sub-eikonal corrections are unlikely to change the $v_2\{\text{SP}\}\approx 0$ result.

Another obvious source of missing relevant physics is the treatment of pre-thermalization time energy loss, which becomes more relevant as the system size decreases. However, despite the increase in the importance of energy loss at early times in small systems, we do not expect the inclusion of early time energy loss to change our prediction that $v_2\{\text{SP}\}\approx 0$ holds robustly across energy loss models, since excluding energy loss before thermalization has been shown to increase the high-$p_T$ $v_2$ \cite{Andres:2019eus,Zigic:2019sth,Stojku:2020wkh,Andres:2022bql}. One can get a qualitative feel for why excluding early time energy loss enhances high-$p_T$ $v_2$ by thinking of drilling a hole of radius $\tau_0$ in the middle of a collision: the presence of this hole enhances the relative size difference between the major and minor axis of the geometry.

A final potentially important physical effect that we will discuss that is missing is an initial state correlation between the geometry and the initial direction of propagation of the hard particle \cite{Dusling:2013oia,Hagiwara:2017ofm,Blok:2017pui,Mace:2018vwq,Soudi:2023epi,JETSCAPE:2024dgu,Soudi:2024slz}. A promising future direction of work is to investigate soft-hard production within one theoretical framework \cite{Carrio:2026xrr}.

Another possible direction of future theoretical work is to investigate to what extent $\psi_2^{\text{hard}}$ and $\psi_n^{\text{soft}}$ are correlated for $n\neq2$. Since the different Fourier modes are orthogonal and independent, it would be surprising if the different planes are correlated; but, if they are correlated, this correlation could provide a powerful experimental method for extracting high-$p_T$ $v_2$.

Let us briefly mention potential future directions of experimental work that may provide insights into the physics of high-$p_T$ anisotropy in small systems. Since the $v_2\{2\}$ measurement suffers from the decorrelation between $\psi_2^{\text{hard}}$ and $\psi_n^{\text{soft}}$ when one of the particles is soft, although statistics limited, a measurement of $v_2\{2\}$ with both particles hard should be very interesting. It would be very interesting to see if the measured $v_2$ depends on quark mass; a $v_2$ measurement insensitive to quark mass would almost certainly be due to physics other final state energy loss. Although difficult, a measurement of electroweak $v_2$, which is inherently unaffected by final state energy loss, would provide a valuable baseline to compare against hadronic high-$p_T$ $v_2$. The striking CMS measurement \cite{CMS:2025kzg} of high-$p_T$ $v_2$ in $p+\mathrm{Pb}$ collisions using four particle correlations with four subevents, which should be less sensitive to non-flow effects, is of a similar size to high-$p_T$ $v_2$ in $\mathrm{Pb}+\mathrm{Pb}$ collisions. This measurement is a factor of four larger than the two-particle correlation measurements of $v_2$ in $p+\mathrm{Pb}$ collisions by ATLAS and ALICE \cite{ATLAS:2019vcm,ALICE:2022cwa}. It will be important for ATLAS and/or ALICE to confirm this CMS result. 

Most interesting would be a measurement of high-$p_T$ anisotropy in $\mathrm{Ne}+\mathrm{Ne}$ and $\mathrm{O}+\mathrm{O}$ collisions. A small high-$p_T$ $v_2$ measurement would provide strong evidence for our picture of final state energy loss as the dominant mechanism for describing the suppression of hard particles in symmetric collision systems. Further, such a measurement of small high-$p_T$ $v_2$ in $\mathrm{Ne}+\mathrm{Ne}$ and $\mathrm{O}+\mathrm{O}$  would provide strong evidence that the large measured high-$p_T$ $v_2$ in $p+\mathrm{Pb}$ collisions is due to experimental and/or physical effects that are unique to small asymmetric collision systems.  On the other hand a measurement of large high-$p_T$ $v_2$ in $\mathrm{Ne}+\mathrm{Ne}$ and $\mathrm{O}+\mathrm{O}$ collisions would suggest that the experimental and/or physical origin(s) of the large high-$p_T$ $v_2$ in $p+\mathrm{Pb}$ collisions are also present in small symmetric collisions.

\section*{Acknowledgements}
The authors would like to thank the South African National Research Foundation (NRF), the National Institute for Theoretical and Computational Sciences (NITheCS), and the SA-CERN collaboration. This research was conducted in part by WAH while visiting the Okinawa Institute of Science and Technology (OIST) through the Theoretical Sciences Visiting Program (TSVP). CF gratefully acknowledges the hospitality of the CERN Theory Group during the completion of this work. Computations were performed using facilities provided by the University of Cape Town's ICTS High Performance Computing team: \href{https://hpc.uct.ac.za}{hpc.uct.ac.za} -- \url{https://doi.org/10.5281/zenodo.10021612}

\bibliographystyle{unsrt}
\bibliography{refs.bib}

\begin{thebibliography}{100}

\bibitem{Harris:1996zx}
John~W. Harris and Berndt Muller.
\newblock {The Search for the quark - gluon plasma}.
\newblock {\em Ann. Rev. Nucl. Part. Sci.}, 46:71--107, 1996.

\bibitem{Niida:2021wut}
T.~Niida and Y.~Miake.
\newblock {Signatures of QGP at RHIC and the LHC}.
\newblock {\em AAPPS Bull.}, 31(1):12, 2021.

\bibitem{Harris:2023tti}
John~W. Harris and Berndt M{\"u}ller.
\newblock {''QGP Signatures'' Revisited}.
\newblock {\em Eur. Phys. J. C}, 84(3):247, 2024.

\bibitem{Pasechnik:2016wkt}
Roman Pasechnik and Michal {\v{S}}umbera.
\newblock {Phenomenological Review on Quark{\textendash}Gluon Plasma: Concepts vs. Observations}.
\newblock {\em Universe}, 3(1):7, 2017.

\bibitem{STAR:2000ekf}
K.~H. Ackermann et~al.
\newblock {Elliptic flow in Au + Au collisions at $\sqrt{s_{NN}}$ = 130 GeV}.
\newblock {\em Phys. Rev. Lett.}, 86:402--407, 2001.

\bibitem{STAR:2002hbo}
C.~Adler et~al.
\newblock {Elliptic flow from two and four particle correlations in Au+Au collisions at $\sqrt{s_{NN}}$ = 130 GeV}.
\newblock {\em Phys. Rev. C}, 66:034904, 2002.

\bibitem{ALICE:2010suc}
K~Aamodt et~al.
\newblock {Elliptic flow of charged particles in Pb-Pb collisions at 2.76 TeV}.
\newblock {\em Phys. Rev. Lett.}, 105:252302, 2010.

\bibitem{Rafelski:1982pu}
Johann Rafelski and Berndt Muller.
\newblock {Strangeness Production in the Quark - Gluon Plasma}.
\newblock {\em Phys. Rev. Lett.}, 48:1066, 1982.
\newblock [Erratum: Phys.Rev.Lett. 56, 2334 (1986)].

\bibitem{STAR:2003jis}
J.~Adams et~al.
\newblock {Multistrange baryon production in Au-Au collisions at $\sqrt{s_{NN}}$ = 130 GeV}.
\newblock {\em Phys. Rev. Lett.}, 92:182301, 2004.

\bibitem{ALICE:2013xmt}
Betty~Bezverkhny Abelev et~al.
\newblock {Multi-strange baryon production at mid-rapidity in Pb-Pb collisions at $\sqrt{s_{NN}}$ = 2.76 TeV}.
\newblock {\em Phys. Lett. B}, 728:216--227, 2014.
\newblock [Erratum: Phys.Lett.B 734, 409--410 (2014)].

\bibitem{Matsui:1986dk}
T.~Matsui and H.~Satz.
\newblock {$J/\psi$ Suppression by Quark-Gluon Plasma Formation}.
\newblock {\em Phys. Lett. B}, 178:416--422, 1986.

\bibitem{CMS:2012bms}
Serguei Chatrchyan et~al.
\newblock {Suppression of Non-Prompt $J/\psi$, Prompt $J/\psi$, and Y(1S) in PbPb Collisions at $\sqrt{s_{NN}}=2.76$ TeV}.
\newblock {\em JHEP}, 05:063, 2012.

\bibitem{ALICE:2012jsl}
Betty Abelev et~al.
\newblock {$J/\psi$ suppression at forward rapidity in Pb-Pb collisions at $\sqrt{s_{NN}}=2.76$ TeV}.
\newblock {\em Phys. Rev. Lett.}, 109:072301, 2012.

\bibitem{Bjorken1982}
J.~D. Bjorken.
\newblock Energy loss of energetic partons in quark-gluon plasma: Possible extinction of high p\_t jets in hadron-hadron collisions.
\newblock Technical Report FERMILAB-Pub-82/59-THY, Fermi National Accelerator Laboratory, Batavia, IL, USA, August 1982.

\bibitem{ALICE:2010yje}
K.~Aamodt et~al.
\newblock {Suppression of Charged Particle Production at Large Transverse Momentum in Central Pb-Pb Collisions at $\sqrt{s_{NN}} =$ 2.76 TeV}.
\newblock {\em Phys. Lett. B}, 696:30--39, 2011.

\bibitem{CMS:2012aa}
Serguei Chatrchyan et~al.
\newblock {Study of High-pT Charged Particle Suppression in PbPb Compared to $pp$ Collisions at $\sqrt{s_{NN}}=2.76$ TeV}.
\newblock {\em Eur. Phys. J. C}, 72:1945, 2012.

\bibitem{ATLAS:2015qmb}
Georges Aad et~al.
\newblock {Measurement of charged-particle spectra in Pb+Pb collisions at $\sqrt{{s}_\mathsf{{NN}}} = 2.76$ TeV with the ATLAS detector at the LHC}.
\newblock {\em JHEP}, 09:050, 2015.

\bibitem{CMS:2016xef}
Vardan Khachatryan et~al.
\newblock {Charged-particle nuclear modification factors in PbPb and pPb collisions at $ \sqrt{s_{\mathrm{N}\;\mathrm{N}}}=5.02 $ TeV}.
\newblock {\em JHEP}, 04:039, 2017.

\bibitem{Vitev:2002pf}
Ivan Vitev and Miklos Gyulassy.
\newblock {High $p_{T}$ tomography of $d$ + Au and Au+Au at SPS, RHIC, and LHC}.
\newblock {\em Phys. Rev. Lett.}, 89:252301, 2002.

\bibitem{Gyulassy:2004zy}
Miklos Gyulassy and Larry McLerran.
\newblock {New forms of QCD matter discovered at RHIC}.
\newblock {\em Nucl. Phys. A}, 750:30--63, 2005.

\bibitem{Wang:2025lct}
Xin-Nian Wang and Urs~Achim Wiedemann.
\newblock {QGP@50: More than Four Decades of Jet Quenching}, 2025.

\bibitem{Betz:2014cza}
Barbara Betz and Miklos Gyulassy.
\newblock {Constraints on the Path-Length Dependence of Jet Quenching in Nuclear Collisions at RHIC and LHC}.
\newblock {\em JHEP}, 08:090, 2014.
\newblock [Erratum: JHEP 10, 043 (2014)].

\bibitem{Xu:2014tda}
Jiechen Xu, Jinfeng Liao, and Miklos Gyulassy.
\newblock {Consistency of Perfect Fluidity and Jet Quenching in semi-Quark-Gluon Monopole Plasmas}.
\newblock {\em Chin. Phys. Lett.}, 32(9):092501, 2015.

\bibitem{Noronha-Hostler:2016eow}
Jacquelyn Noronha-Hostler, Barbara Betz, Jorge Noronha, and Miklos Gyulassy.
\newblock {Event-by-event hydrodynamics $+$ jet energy loss: A solution to the $R_{AA} \otimes v_2$ puzzle}.
\newblock {\em Phys. Rev. Lett.}, 116(25):252301, 2016.

\bibitem{Zigic:2018ovr}
Dusan Zigic, Igor Salom, Jussi Auvinen, Marko Djordjevic, and Magdalena Djordjevic.
\newblock {DREENA-B framework: first predictions of $R_{AA}$ and $v_2$ within dynamical energy loss formalism in evolving QCD medium}.
\newblock {\em Phys. Lett. B}, 791:236--241, 2019.

\bibitem{Xu:2015bbz}
Jiechen Xu, Jinfeng Liao, and Miklos Gyulassy.
\newblock {Bridging Soft-Hard Transport Properties of Quark-Gluon Plasmas with CUJET3.0}.
\newblock {\em JHEP}, 02:169, 2016.

\bibitem{ATLAS:2012cix}
Georges Aad et~al.
\newblock {Observation of Associated Near-Side and Away-Side Long-Range Correlations in $\sqrt{s_{NN}}$=5.02 TeV Proton-Lead Collisions with the ATLAS Detector}.
\newblock {\em Phys. Rev. Lett.}, 110(18):182302, 2013.

\bibitem{ATLAS:2013jmi}
Georges Aad et~al.
\newblock {Measurement with the ATLAS detector of multi-particle azimuthal correlations in p+Pb collisions at $\sqrt{s_{NN}}$ =5.02 TeV}.
\newblock {\em Phys. Lett. B}, 725:60--78, 2013.

\bibitem{ALICE:2014dwt}
Betty~Bezverkhny Abelev et~al.
\newblock {Multiparticle azimuthal correlations in p -Pb and Pb-Pb collisions at the CERN Large Hadron Collider}.
\newblock {\em Phys. Rev. C}, 90(5):054901, 2014.

\bibitem{CMS:2015yux}
Vardan Khachatryan et~al.
\newblock {Evidence for Collective Multiparticle Correlations in p-Pb Collisions}.
\newblock {\em Phys. Rev. Lett.}, 115(1):012301, 2015.

\bibitem{ALICE:2013wgn}
Betty~Bezverkhny Abelev et~al.
\newblock {Multiplicity Dependence of Pion, Kaon, Proton and Lambda Production in p-Pb Collisions at $\sqrt{s_{NN}}$ = 5.02 TeV}.
\newblock {\em Phys. Lett. B}, 728:25--38, 2014.

\bibitem{ALICE:2015mpp}
Jaroslav Adam et~al.
\newblock {Multi-strange baryon production in p-Pb collisions at $\sqrt{s_{NN}}=5.02$ TeV}.
\newblock {\em Phys. Lett. B}, 758:389--401, 2016.

\bibitem{ALICE:2016sdt}
Jaroslav Adam et~al.
\newblock {Centrality dependence of $\mathbf{\psi}$(2S) suppression in p-Pb collisions at $\mathbf{\sqrt{{\textit s}_{\rm NN}}}$ = 5.02 TeV}.
\newblock {\em JHEP}, 06:050, 2016.

\bibitem{ATLAS:2022kqu}
Georges Aad et~al.
\newblock {Charged-hadron production in $pp$, $p$+Pb, Pb+Pb, and Xe+Xe collisions at $\sqrt{s_{_\text{NN}}}=5$ TeV with the ATLAS detector at the LHC}.
\newblock {\em JHEP}, 07:074, 2023.

\bibitem{ALICE:2021est}
Shreyasi Acharya et~al.
\newblock {Nuclear modification factor of light neutral-meson spectra up to high transverse momentum in p{\textendash}Pb collisions at sNN=8.16 TeV}.
\newblock {\em Phys. Lett. B}, 827:136943, 2022.

\bibitem{FH2025-2}
Coleridge Faraday and W.~A. Horowitz.
\newblock {A unified description of small, peripheral, and large system suppression data from pQCD}.
\newblock {\em Phys. Lett. B}, 864:139437, 2025.

\bibitem{Zhang:2013oca}
Xilin Zhang and Jinfeng Liao.
\newblock {Jet Quenching and Its Azimuthal Anisotropy in AA and possibly High Multiplicity pA and dA Collisions}.
\newblock arXiv:1311.5463, 2013.
\newblock [nucl-th].

\bibitem{Kang:2015mta}
Zhong-Bo Kang, Ivan Vitev, and Hongxi Xing.
\newblock {Effects of cold nuclear matter energy loss on inclusive jet production in p+A collisions at energies available at the BNL Relativistic Heavy Ion Collider and the CERN Large Hadron Collider}.
\newblock {\em Phys. Rev. C}, 92(5):054911, 2015.

\bibitem{Ke:2022gkq}
Weiyao Ke and Ivan Vitev.
\newblock {Searching for QGP droplets with high-pT hadrons and heavy flavor}.
\newblock {\em Phys. Rev. C}, 107(6):064903, 2023.

\bibitem{PHENIX:2023dxl}
N.~J. Abdulameer et~al.
\newblock {Disentangling Centrality Bias and Final-State Effects in the Production of High-pT Neutral Pions Using Direct Photon in d+Au Collisions at sNN=200{\,}{\,}GeV}.
\newblock {\em Phys. Rev. Lett.}, 134(2):022302, 2025.

\bibitem{ATLAS:2019vcm}
Georges Aad et~al.
\newblock {Transverse momentum and process dependent azimuthal anisotropies in $\sqrt{s_{\mathrm{NN}}}=8.16$ TeV $p$+Pb collisions with the ATLAS detector}.
\newblock {\em Eur. Phys. J. C}, 80(1):73, 2020.

\bibitem{ALICE:2022cwa}
Shreyasi Acharya et~al.
\newblock {Azimuthal anisotropy of jet particles in p-Pb and Pb-Pb collisions at $ \sqrt{s_{NN}} $ = 5.02 TeV}.
\newblock {\em JHEP}, 08:234, 2024.

\bibitem{CMS:2025kzg}
Vladimir Chekhovsky et~al.
\newblock {Evidence for Similar Collectivity of High Transverse-Momentum Particles in p-Pb and Pb-Pb Collisions}.
\newblock {\em Phys. Rev. Lett.}, 135(7):071903, 2025.

\bibitem{CMS:2025bta}
Aram Hayrapetyan et~al.
\newblock {Discovery of suppressed charged-particle production in ultrarelativistic oxygen-oxygen collisions}.
\newblock 10 2025.

\bibitem{ALICE:2025oop}
{ALICE Collaboration}.
\newblock {ALI-PREL-617876}.
\newblock \url{https://alice-figure.web.cern.ch/node/35718}, 2024.

\bibitem{vanderSchee:2025hoe}
Wilke van~der Schee, Isobel Kolb{\'e}, Govert Nijs, Kumail Ruhani, Ishtiaq Ahmed, and Shahin Iqbal.
\newblock {Three models for charged hadron nuclear modification from light to heavy ions}, 9 2025.

\bibitem{Pablos:2025cli}
Daniel Pablos and Adam Takacs.
\newblock {Bayesian Constraints on Pre-Equilibrium Jet Quenching and Predictions for Oxygen Collisions}, 9 2025.

\bibitem{Zakharov:2025mbk}
B.~G. Zakharov.
\newblock {Predictions for RAA in 5.36 TeV C + C, O + O, and Ne + Ne collisions at the LHC}.
\newblock {\em Pisma Zh. Eksp. Teor. Fiz.}, 122(8):437--438, 2025.

\bibitem{Mazeliauskas:2025clt}
Aleksas Mazeliauskas.
\newblock {Energy loss baseline for light hadrons in oxygen-oxygen collisions at $\sqrt{s_\mathrm{NN}}=5.36\,\text{TeV}$}, 9 2025.

\bibitem{FBBVW}
Coleridge Faraday, Ben Bert, Jack Brand, Werner Vogelsang, and W.~A. Horowitz.
\newblock {From Lead to Helium: Discovery Potential for Jet Quenching in the Smallest Collision Systems}.
\newblock 12 2025.

\bibitem{ALICE:2018lyv}
S.~Acharya et~al.
\newblock {Measurement of D$^{0}$, D$^{+}$, D$^{*+}$ and D$_{s}^{+}$ production in Pb-Pb collisions at $ \sqrt{{\mathrm{s}}_{\mathrm{NN}}}=5.02 $ TeV}.
\newblock {\em JHEP}, 10:174, 2018.

\bibitem{ALICE:2018vuu}
S.~Acharya et~al.
\newblock {Transverse momentum spectra and nuclear modification factors of charged particles in pp, p-Pb and Pb-Pb collisions at the LHC}.
\newblock {\em JHEP}, 11:013, 2018.

\bibitem{ALICE:2019hno}
Shreyasi Acharya et~al.
\newblock {Production of charged pions, kaons, and (anti-)protons in Pb-Pb and inelastic $pp$ collisions at $\sqrt {s_{NN}}$ = 5.02 TeV}.
\newblock {\em Phys. Rev. C}, 101(4):044907, 2020.

\bibitem{CMS:2017qjw}
Albert~M Sirunyan et~al.
\newblock {Nuclear modification factor of D$^0$ mesons in PbPb collisions at $\sqrt{s_\mathrm{NN}} = 5.02$ TeV}.
\newblock {\em Phys. Lett. B}, 782:474--496, 2018.

\bibitem{ALICE:2017pbx}
Shreyasi Acharya et~al.
\newblock {$D$-meson azimuthal anisotropy in midcentral Pb-Pb collisions at ${\sqrt{s_{NN}}}$ = 5.02 TeV}.
\newblock {\em Phys. Rev. Lett.}, 120(10):102301, 2018.

\bibitem{ALICE:2018rtz}
S.~Acharya et~al.
\newblock {Energy dependence and fluctuations of anisotropic flow in Pb-Pb collisions at $ \sqrt{s_{NN}}=5.02 $ and 2.76 TeV}.
\newblock {\em JHEP}, 07:103, 2018.

\bibitem{ATLAS:2018ezv}
Morad Aaboud et~al.
\newblock {Measurement of the azimuthal anisotropy of charged particles produced in $\sqrt{s_{NN}}$ = 5.02 TeV Pb+Pb collisions with the ATLAS detector}.
\newblock {\em Eur. Phys. J. C}, 78(12):997, 2018.

\bibitem{CMS:2017vhp}
Albert~M Sirunyan et~al.
\newblock {Measurement of prompt $D^0$ meson azimuthal anisotropy in Pb-Pb collisions at $\sqrt{{s}_{NN}}$ = 5.02 TeV}.
\newblock {\em Phys. Rev. Lett.}, 120(20):202301, 2018.

\bibitem{CMS:2017xgk}
A.~M. Sirunyan et~al.
\newblock {Azimuthal anisotropy of charged particles with transverse momentum up to 100 GeV/ c in PbPb collisions at $\sqrt {s}_{{NN}}$=5.02 TeV}.
\newblock {\em Phys. Lett. B}, 776:195--216, 2018.

\bibitem{CMS:2020bnz}
Albert~M Sirunyan et~al.
\newblock {Measurement of prompt ${\mathrm{D^0}}$ and ${\mathrm{\overline{D}}{}^0}$ meson azimuthal anisotropy and search for strong electric fields in PbPb collisions at $\sqrt{s_\mathrm{NN}} =$ 5.02 TeV}.
\newblock {\em Phys. Lett. B}, 816:136253, 2021.

\bibitem{CMS:2022vfn}
Armen Tumasyan et~al.
\newblock {Measurements of azimuthal anisotropy of nonprompt D0 mesons in PbPb collisions at sNN=5.02TeV}.
\newblock {\em Phys. Lett. B}, 850:138389, 2024.

\bibitem{FH2025-3}
Coleridge Faraday and W.~A. Horowitz.
\newblock {Statistical analysis of pQCD energy loss across system size, flavor, $ \sqrt{s_{NN}} $, and p$_{T}$}.
\newblock {\em JHEP}, 11:019, 2025.

\bibitem{WHDG2007}
Simon Wicks, William Horowitz, Magdalena Djordjevic, and Miklos Gyulassy.
\newblock {Elastic, inelastic, and path length fluctuations in jet tomography}.
\newblock {\em Nucl. Phys. A}, 784:426--442, 2007.

\bibitem{DGLV1}
M.~Gyulassy, P.~Levai, and I.~Vitev.
\newblock {Reaction operator approach to nonAbelian energy loss}.
\newblock {\em Nucl. Phys. B}, 594:371--419, 2001.

\bibitem{DGLV2}
Miklos Gyulassy, Peter Levai, and Ivan Vitev.
\newblock {Jet quenching in thin quark gluon plasmas. 1. Formalism}.
\newblock {\em Nucl. Phys. B}, 571:197--233, 2000.

\bibitem{DGLV3}
Magdalena Djordjevic and Miklos Gyulassy.
\newblock {Heavy quark radiative energy loss in QCD matter}.
\newblock {\em Nucl. Phys. A}, 733:265--298, 2004.

\bibitem{Kolbe:MSc}
Isobel Kolbe.
\newblock {Short path length pQCD corrections to energy loss in the quark gluon plasma}.
\newblock Master's thesis, Cape Town U., 2015.

\bibitem{Kolbe:SPLC}
Isobel Kolbe and W.~A. Horowitz.
\newblock {Short path length corrections to Djordjevic-Gyulassy-Levai-Vitev energy loss}.
\newblock {\em Phys. Rev. C}, 100(2):024913, 2019.

\bibitem{HTL1}
H.~Arthur Weldon.
\newblock {Covariant Calculations at Finite Temperature: The Relativistic Plasma}.
\newblock {\em Phys. Rev. D}, 26:1394, 1982.

\bibitem{HTL5}
H.~Arthur Weldon.
\newblock {Effective Fermion Masses of Order gT in High Temperature Gauge Theories with Exact Chiral Invariance}.
\newblock {\em Phys. Rev. D}, 26:2789, 1982.

\bibitem{HTL2}
Robert~D. Pisarski.
\newblock {Scattering Amplitudes in Hot Gauge Theories}.
\newblock {\em Phys. Rev. Lett.}, 63:1129, 1989.

\bibitem{HTL3}
V.~V. Klimov.
\newblock {Collective Excitations in a Hot Quark Gluon Plasma}.
\newblock {\em Sov. Phys. JETP}, 55:199--204, 1982.

\bibitem{HTL4}
Eric Braaten and Robert~D. Pisarski.
\newblock {Soft Amplitudes in Hot Gauge Theories: A General Analysis}.
\newblock {\em Nucl. Phys. B}, 337:569--634, 1990.

\bibitem{BT1}
Eric Braaten and Markus~H. Thoma.
\newblock {Energy loss of a heavy fermion in a hot plasma}.
\newblock {\em Phys. Rev. D}, 44:1298--1310, 1991.

\bibitem{BT2}
Eric Braaten and Markus~H. Thoma.
\newblock {Energy loss of a heavy quark in the quark - gluon plasma}.
\newblock {\em Phys. Rev. D}, 44(9):R2625, 1991.

\bibitem{WicksPHD}
Simon Wicks.
\newblock {Fluctuations with small numbers: Developing the perturbative paradigm for jet physics in the QGP at RHIC and LHC}.
\newblock Other thesis, Columbia University, 2008.

\bibitem{FH2023}
Coleridge Faraday, Antonia Grindrod, and W.~A. Horowitz.
\newblock {Inconsistencies in, and short pathlength correction to, $R_{AA}(p_T)$ in~$\textrm{A}+\textrm{A}$ and $\textrm{p} + \textrm{A}$ collisions}.
\newblock {\em Eur. Phys. J. C}, 83(11):1060, 2023.

\bibitem{FH2025-1}
Coleridge Faraday and W.~A. Horowitz.
\newblock {Collisional and radiative energy loss in small systems}.
\newblock {\em Phys. Rev. C}, 111(5):054911, 2025.

\bibitem{FH2023SAIP}
Coleridge Faraday and W.~A. Horowitz.
\newblock {Assumption Breakdown in Radiative Energy Loss}.
\newblock In {\em {67th Annual Conference of the South African Institute of Physics}}, 9 2023.

\bibitem{Andres:2023jao}
Carlota Andres, Liliana Apolin{\'a}rio, Fabio Dominguez, and Marcos~Gonzalez Martinez.
\newblock {In-medium gluon radiation spectrum with all-order resummation of multiple scatterings in longitudinally evolving media}.
\newblock {\em JHEP}, 11:025, 2024.

\bibitem{Weisstein2025LevenbergMarquardt}
Eric~W. Weisstein.
\newblock Levenberg--marquardt method.
\newblock MathWorld -- A Wolfram Resource, 2025.
\newblock Accessed: 2025-12-26.

\bibitem{Wang:2003aw}
Xin-Nian Wang.
\newblock {Why the observed jet quenching at RHIC is due to parton energy loss}.
\newblock {\em Phys. Lett. B}, 579:299--308, 2004.

\bibitem{Elfner:2020men}
Hannah Elfner, Philipp Dorau, Jean-Bernard Rose, and Daniel Pablos.
\newblock {Jet quenching in the hadron gas: an exploratory study}.
\newblock {\em PoS}, HardProbes2020:155, 2021.

\bibitem{Ipp:2020nfu}
Andreas Ipp, David~I. M{\"u}ller, and Daniel Schuh.
\newblock {Jet momentum broadening in the pre-equilibrium Glasma}.
\newblock {\em Phys. Lett. B}, 810:135810, 2020.

\bibitem{Andres:2022bql}
Carlota Andres, Liliana Apolin{\'a}rio, Fabio Dominguez, Marcos~Gonzalez Martinez, and Carlos~A. Salgado.
\newblock {Medium-induced radiation with vacuum propagation in the pre-hydrodynamics phase}.
\newblock {\em JHEP}, 03:189, 2023.

\bibitem{Avramescu:2023qvv}
Dana Avramescu, Virgil B{\u{a}}ran, Vincenzo Greco, Andreas Ipp, David.~I. M{\"u}ller, and Marco Ruggieri.
\newblock {Simulating jets and heavy quarks in the glasma using the colored particle-in-cell method}.
\newblock {\em Phys. Rev. D}, 107(11):114021, 2023.

\bibitem{Barata:2024xwy}
Jo{\~a}o Barata, Sigtryggur Hauksson, Xo{\'a}n Mayo~L{\'o}pez, and Andrey~V. Sadofyev.
\newblock {Jet quenching in the glasma phase: Medium-induced radiation}.
\newblock {\em Phys. Rev. D}, 110(9):094055, 2024.

\bibitem{ShenPrivateComm}
C.~Shen, 2025.
\newblock private communication.

\bibitem{IP-Glasma1}
Bjoern Schenke, Chun Shen, and Prithwish Tribedy.
\newblock {Running the gamut of high energy nuclear collisions}.
\newblock {\em Phys. Rev. C}, 102(4):044905, 2020.

\bibitem{IP-Glasma2}
Bjoern Schenke, Prithwish Tribedy, and Raju Venugopalan.
\newblock {Event-by-event gluon multiplicity, energy density, and eccentricities in ultrarelativistic heavy-ion collisions}.
\newblock {\em Phys. Rev. C}, 86:034908, 2012.

\bibitem{IP-Glasma3}
Bjoern Schenke, Prithwish Tribedy, and Raju Venugopalan.
\newblock {Fluctuating Glasma initial conditions and flow in heavy ion collisions}.
\newblock {\em Phys. Rev. Lett.}, 108:252301, 2012.

\bibitem{MUSIC1}
Bjorn Schenke, Sangyong Jeon, and Charles Gale.
\newblock {Elliptic and triangular flow in event-by-event (3+1)D viscous hydrodynamics}.
\newblock {\em Phys. Rev. Lett.}, 106:042301, 2011.

\bibitem{MUSIC2}
Bjorn Schenke, Sangyong Jeon, and Charles Gale.
\newblock {Higher flow harmonics from (3+1)D event-by-event viscous hydrodynamics}.
\newblock {\em Phys. Rev. C}, 85:024901, 2012.

\bibitem{MUSIC3}
Bjoern Schenke, Sangyong Jeon, and Charles Gale.
\newblock {(3+1)D hydrodynamic simulation of relativistic heavy-ion collisions}.
\newblock {\em Phys. Rev. C}, 82:014903, 2010.

\bibitem{BertFaradayHorowitz2026}
Ben Bert, Coleridge Faraday, and W.~A. Horowitz.
\newblock {Work in preparation}, 2026.

\bibitem{Moreland:2014oya}
J.~Scott Moreland, Jonah~E. Bernhard, and Steffen~A. Bass.
\newblock {Alternative ansatz to wounded nucleon and binary collision scaling in high-energy nuclear collisions}.
\newblock {\em Phys. Rev. C}, 92(1):011901, 2015.

\bibitem{Werner:2010aa}
K.~Werner, Iu. Karpenko, T.~Pierog, M.~Bleicher, and K.~Mikhailov.
\newblock {Event-by-Event Simulation of the Three-Dimensional Hydrodynamic Evolution from Flux Tube Initial Conditions in Ultrarelativistic Heavy Ion Collisions}.
\newblock {\em Phys. Rev. C}, 82:044904, 2010.

\bibitem{Shen:2014vra}
Chun Shen, Zhi Qiu, Huichao Song, Jonah Bernhard, Steffen Bass, and Ulrich Heinz.
\newblock {The iEBE-VISHNU code package for relativistic heavy-ion collisions}.
\newblock {\em Comput. Phys. Commun.}, 199:61--85, 2016.

\bibitem{ALICE:2022wpn}
Shreyasi Acharya et~al.
\newblock {The ALICE experiment: a journey through QCD}.
\newblock {\em Eur. Phys. J. C}, 84(8):813, 2024.

\bibitem{CMS:2024krd}
Aram Hayrapetyan et~al.
\newblock {Overview of high-density QCD studies with the CMS experiment at the LHC}.
\newblock {\em Phys. Rept.}, 1115:219--367, 2025.

\bibitem{PHENIX:2004vcz}
K.~Adcox et~al.
\newblock {Formation of dense partonic matter in relativistic nucleus-nucleus collisions at RHIC: Experimental evaluation by the PHENIX collaboration}.
\newblock {\em Nucl. Phys. A}, 757:184--283, 2005.

\bibitem{STAR:2005gfr}
John Adams et~al.
\newblock {Experimental and theoretical challenges in the search for the quark gluon plasma: The STAR Collaboration's critical assessment of the evidence from RHIC collisions}.
\newblock {\em Nucl. Phys. A}, 757:102--183, 2005.

\bibitem{ALICE:2018ekf}
Shreyasi Acharya et~al.
\newblock {Analysis of the apparent nuclear modification in peripheral Pb{\textendash}Pb collisions at 5.02 TeV}.
\newblock {\em Phys. Lett. B}, 793:420--432, 2019.

\bibitem{CMS:2026qef}
Andrey Belyaev et~al.
\newblock {System-size dependence of charged-particle suppression in ultrarelativistic nucleus-nucleus collisions}.
\newblock 2 2026.

\bibitem{Horowitz:2009eb}
W.~A. Horowitz and B.~A. Cole.
\newblock {Systematic theoretical uncertainties in jet quenching due to gluon kinematics}.
\newblock {\em Phys. Rev. C}, 81:024909, 2010.

\bibitem{EfronTibshirani1993}
Bradley Efron and Robert~J. Tibshirani.
\newblock {\em An Introduction to the Bootstrap}.
\newblock Chapman \& Hall, New York, 1993.

\bibitem{Zigic:2019sth}
Dusan Zigic, Bojana Ilic, Marko Djordjevic, and Magdalena Djordjevic.
\newblock {Exploring the initial stages in heavy-ion collisions with high-$p_\bot$ $R_{AA}$ and $v_2$ theory and data}.
\newblock {\em Phys. Rev. C}, 101(6):064909, 2020.

\bibitem{Andres:2019eus}
Carlota Andres, N{\'e}stor Armesto, Harri Niemi, Risto Paatelainen, and Carlos~A. Salgado.
\newblock {Jet quenching as a probe of the initial stages in heavy-ion collisions}.
\newblock {\em Phys. Lett. B}, 803:135318, 2020.

\bibitem{Dorau:2019ozd}
Philipp Dorau, Jean-Bernard Rose, Daniel Pablos, and Hannah Elfner.
\newblock {Jet Quenching in the Hadron Gas: An Exploratory Study}.
\newblock {\em Phys. Rev. C}, 101(3):035208, 2020.

\bibitem{Datta:2025gql}
Ritoban Datta and Abhijit Majumder.
\newblock {The Effect of Hadronic Matter on Parton Energy Loss}, 12 2025.

\bibitem{Buzzatti:2012dy}
Alessandro Buzzatti and Miklos Gyulassy.
\newblock {A running coupling explanation of the surprising transparency of the QGP at LHC}.
\newblock {\em Nucl. Phys. A}, 904-905:779c--782c, 2013.

\bibitem{Zakharov:2018cpv}
B.~G. Zakharov.
\newblock {On the Use of the Running Coupling Constant $\alpha_{s}$ in Calculations of Radiative Energy Losses of Fast Partons in a Quark{\textendash}Gluon Plasma}.
\newblock {\em JETP Lett.}, 107(2):73--78, 2018.

\bibitem{Luzum:2013yya}
Matthew Luzum and Hannah Petersen.
\newblock {Initial State Fluctuations and Final State Correlations in Relativistic Heavy-Ion Collisions}.
\newblock {\em J. Phys. G}, 41:063102, 2014.

\bibitem{Snellings:2011sz}
Raimond Snellings.
\newblock {Elliptic Flow: A Brief Review}.
\newblock {\em New J. Phys.}, 13:055008, 2011.

\bibitem{ATLAS:2018ofq}
Morad Aaboud et~al.
\newblock {Measurement of the suppression and azimuthal anisotropy of muons from heavy-flavor decays in Pb+Pb collisions at $\sqrt{s_{\mathrm{NN}}} = 2.76$ TeV with the ATLAS detector}.
\newblock {\em Phys. Rev. C}, 98(4):044905, 2018.

\bibitem{ATLAS:2024mch}
Georges Aad et~al.
\newblock {Azimuthal anisotropies of charged particles with high transverse momentum in Pb+Pb collisions at $\sqrt{s_{_\text{NN}}} = 5.02$ TeV with the ATLAS detector}.
\newblock {\em Phys. Rev. C}, 112(2):024910, 2025.

\bibitem{ALICE:2014qvj}
Betty~Bezverkhny Abelev et~al.
\newblock {Azimuthal anisotropy of D meson production in Pb-Pb collisions at $\sqrt{s_{\rm NN}} = 2.76$ TeV}.
\newblock {\em Phys. Rev. C}, 90(3):034904, 2014.

\bibitem{ALICE:2014xsp}
Jaroslav Adam et~al.
\newblock {Centrality dependence of particle production in p-Pb collisions at $\sqrt{s_{\rm NN} }$= 5.02 TeV}.
\newblock {\em Phys. Rev. C}, 91(6):064905, 2015.

\bibitem{Alvioli:2013vk}
M.~Alvioli and M.~Strikman.
\newblock {Color fluctuation effects in proton-nucleus collisions}.
\newblock {\em Phys. Lett. B}, 722:347--354, 2013.

\bibitem{Kordell:2016njg}
Michael Kordell and Abhijit Majumder.
\newblock {Jets in d(p)-A Collisions: Color Transparency or Energy Conservation}.
\newblock {\em Phys. Rev. C}, 97(5):054904, 2018.

\bibitem{Perepelitsa:2024eik}
Dennis~V. Perepelitsa.
\newblock {Contribution to differential {\ensuremath{\pi}}0 and {\ensuremath{\gamma}}dir modification in small systems from color fluctuation effects}.
\newblock {\em Phys. Rev. C}, 110(1):L011901, 2024.

\bibitem{JETSCAPE:2024dgu}
I.~Soudi et~al.
\newblock {Soft-hard framework with exact four-momentum conservation for small systems}.
\newblock {\em Phys. Rev. C}, 112(1):014905, 2025.

\bibitem{Armesto:2004pt}
Nestor Armesto, Carlos~A. Salgado, and Urs~Achim Wiedemann.
\newblock {Measuring the collective flow with jets}.
\newblock {\em Phys. Rev. Lett.}, 93:242301, 2004.

\bibitem{Sadofyev:2021ohn}
Andrey~V. Sadofyev, Matthew~D. Sievert, and Ivan Vitev.
\newblock {Ab~initio coupling of jets to collective flow in the opacity expansion approach}.
\newblock {\em Phys. Rev. D}, 104(9):094044, 2021.

\bibitem{Fu:2022idl}
Yu~Fu, Jorge Casalderrey-Solana, and Xin-Nian Wang.
\newblock {Asymmetric transverse momentum broadening in an inhomogeneous medium}.
\newblock {\em Phys. Rev. D}, 107(5):054038, 2023.

\bibitem{Bahder:2024jpa}
Joseph Bahder, Hasan Rahman, Matthew~D. Sievert, and Ivan Vitev.
\newblock {Signatures of jet drift in quark-gluon plasma hard-probe observables}.
\newblock {\em Phys. Rev. Res.}, 8(1):L012016, 2026.

\bibitem{Stojku:2020wkh}
Stefan Stojku, Jussi Auvinen, Marko Djordjevic, Pasi Huovinen, and Magdalena Djordjevic.
\newblock {Early evolution constrained by high-p{\ensuremath{\perp}} quark-gluon plasma tomography}.
\newblock {\em Phys. Rev. C}, 105(2):L021901, 2022.

\bibitem{Dusling:2013oia}
Kevin Dusling and Raju Venugopalan.
\newblock {Comparison of the color glass condensate to dihadron correlations in proton-proton and proton-nucleus collisions}.
\newblock {\em Phys. Rev. D}, 87(9):094034, 2013.

\bibitem{Hagiwara:2017ofm}
Yoshikazu Hagiwara, Yoshitaka Hatta, Bo-Wen Xiao, and Feng Yuan.
\newblock {Elliptic Flow in Small Systems due to Elliptic Gluon Distributions?}
\newblock {\em Phys. Lett. B}, 771:374--378, 2017.

\bibitem{Blok:2017pui}
Boris Blok, Christian~D. J{\"a}kel, Mark Strikman, and Urs~Achim Wiedemann.
\newblock {Collectivity from interference}.
\newblock {\em JHEP}, 12:074, 2017.

\bibitem{Mace:2018vwq}
Mark Mace, Vladimir~V. Skokov, Prithwish Tribedy, and Raju Venugopalan.
\newblock {Hierarchy of Azimuthal Anisotropy Harmonics in Collisions of Small Systems from the Color Glass Condensate}.
\newblock {\em Phys. Rev. Lett.}, 121(5):052301, 2018.
\newblock [Erratum: Phys.Rev.Lett. 123, 039901 (2019)].

\bibitem{Soudi:2023epi}
Ismail Soudi and Abhijit Majumder.
\newblock {Azimuthal anisotropy at high transverse momentum in p-p and p-A collisions}.
\newblock {\em Phys. Lett. B}, 859:139105, 2024.

\bibitem{Soudi:2024slz}
Ismail Soudi and Abhijit Majumder.
\newblock {T-odd parton distribution functions and azimuthal anisotropy at high transverse momentum in p-p and p-A collisions}.
\newblock {\em Phys. Rev. C}, 111(2):024901, 2025.

\bibitem{Carrio:2026xrr}
Erik Carri{\'o} and Daniel Pablos.
\newblock {Elliptic Anisotropy from Quantum Diffraction}.
\newblock 3 2026.

\end{thebibliography}

\end{document}